\documentclass{aa}

\usepackage{txfonts} % REMOVE newtxtext,newtxmath}
\usepackage[T1]{fontenc}
\usepackage{graphicx}
\usepackage{subcaption}
\usepackage{booktabs}
\usepackage{amsmath}
\usepackage{lipsum}
\usepackage{comment}
\usepackage{adjustbox}
\usepackage{multirow}
\usepackage{txfonts}
\usepackage{float}
\usepackage{natbib}
\usepackage[colorlinks=true, linkcolor=blue, citecolor=blue, urlcolor=blue]{hyperref}
\usepackage{lscape}
\usepackage{placeins}
\usepackage{textgreek}
\usepackage{xspace}
\usepackage{xargs}
\usepackage[usenames,dvipsnames]{xcolor}
\usepackage{enumitem}
\usepackage{pifont}

\makeatletter
\renewcommand*\aa@pageof{, page \thepage{} of \pageref*{LastPage}}
\makeatother

\newcommand{\ha}{\textup{H$\alpha$}\xspace}
\newcommand{\hb}{\textup{H$\beta$}\xspace}
\newcommand{\hg}{\textup{H$\gamma$}\xspace}
\newcommandx{\oiperm}[1][1=$\lambda$8446]{\textup{O\,\textsc{i}\,}{#1}\xspace}
\newcommandx{\oi}[1][1=]{\textup{[O\,\textsc{i}]}{#1}\xspace}
\newcommandx{\oii}[1][1=]{\textup{[O\,\textsc{ii}]}{#1}\xspace}
\newcommandx{\oiii}[1][1=]{\textup{[O\,\textsc{iii}]}{#1}\xspace}
\newcommandx{\ovi}[1][1=]{\textup{O\,\textsc{vi}}{#1}\xspace}
\newcommandx{\nii}[1][1=]{\textup{[N\,\textsc{ii}]}{#1}\xspace}
\newcommandx{\sii}[1][1=]{\textup{[S\,\textsc{ii}]}{#1}\xspace}
\newcommandx{\nai}[1][1=]{\textup{Na\,\textsc{i}}{#1}\xspace}

\newcommandx{\feiiperm}[1][1=]{\textup{Fe\,\textsc{ii}}{#1}\xspace}
\newcommandx{\feii}[1][1=]{\textup{[Fe\,\textsc{ii}]}{#1}\xspace}
\newcommandx{\fevii}[1][1=]{\textup{[Fe\,\textsc{vii}]}{#1}\xspace}
\newcommandx{\fex}[1][1=]{\textup{[Fe\,\textsc{x}]}{#1}\xspace}

\newcommandx{\hei}[1][1=]{\textup{He\,\textsc{i}}{#1}\xspace}
\newcommandx{\heii}[1][1=]{\textup{He\,\textsc{ii}}{#1}\xspace}
\newcommandx{\ariii}[1][1=]{\textup{[Ar\,\textsc{iii}]}{#1}\xspace}
\newcommandx{\ariv}[1][1=]{\textup{[Ar\,\textsc{iv}]}{#1}\xspace}

\newcommandx{\mgii}[1][1=]{\textup{Mg\,\textsc{ii}}{#1}\xspace}
\newcommandx{\niii}[1][1=]{\textup{N\,\textsc{iii}]}{#1}\xspace}
\newcommandx{\niv}[1][1=]{\textup{N\,\textsc{iv}]}{#1}\xspace}
\newcommandx{\nv}[1][1=]{\textup{N\,\textsc{v}}{#1}\xspace}
\newcommandx{\ciii}[1][1=]{\textup{C\,\textsc{iii}]}{#1}\xspace}
\newcommandx{\civ}[1][1=]{\textup{C\,\textsc{iv}}{#1}\xspace}

\newcommand{\mum}{\text{\textmu m}\xspace}
\newcommand{\cmark}{\color{green} \ding{51}} % ✓
\newcommand{\xmark}{\color{red} \ding{55}} % ✗

\newcommand{\fde}[1]{{\color{BrickRed}[FDE:  #1]}}

\usepackage{graphicx}	% Including figure files
\usepackage{amsmath}	% Advanced maths commands

\begin{document}

% Title of the paper, and the short title which is used in the headers.
% Keep the title short and informative.
\title{The Little Blue and Red Dots Rosetta Stones}
\subtitle{Non-Gaussian broad lines, hot dust, and X-ray weakness}
%\title[JWST Little Blue Dots]{Little and Blue, yet peculiar too: a comparative analysis between JWST's brightest Little Red Dot and Little Blue Dot}

% The list of authors, and the short list which is used in the headers.
% If you need two or more lines of authors, add an extra line using \newauthor
\author{
    M. Brazzini \inst{1,2,3,4}
    \and F. D'Eugenio \inst{3,4}
     \and R. Maiolino \inst{3,4,5}
     \and J. Lyu \inst{6}
     \and C. DeCoursey \inst{6}
     \and H.~\"Ubler \inst{7}
     \and X. Ji \inst{3,4}
     \and I. Juod{\v z}balis \inst{3,4}
     \and J. Scholtz \inst{3,4}
     \and G. C. Jones \inst{3,4}
     \and K. Hainline \inst{6}
     \and E. Dalla~Bont\`a \inst{8,9,10}
     \and P. G. P\'erez-Gonz\'alez \inst{11}
     \and S. Geris \inst{3,4}
     \and A. Harshan \inst{3,4}
     \and C. Feruglio \inst{2,12}
     \and M. Bischetti \inst{1,2}
     \and G. Mazzolari \inst{7}
     \and G. Rieke \inst{6}
     \and S. Alberts \inst{6}
     \and B. Trefoloni \inst{13,14}
     \and S. Carniani \inst{13}
     \and E. Parlanti \inst{13}
     \and A. Marconi \inst{14,15}
     \and G. Risaliti \inst{14,15}
     \and C. Ramos Almeida \inst{16,17}
     \and P. Rinaldi \inst{6}
     \and M. Perna \inst{11}
     \and S. Zamora \inst{13}
     \and I. Lamperti \inst{11,14,15}
     \and G. Venturi \inst{13}
     \and G. Cresci \inst{14}
     \and A. J. Bunker \inst{19}
     \and L. R. Ivey \inst{3, 4}
}
\institute{
    Department of Physics, Astronomy Section, University of Trieste, Via G.B. Tiepolo, 11, I-34143 Trieste, Italy
    \and
    INAF - Osservatorio Astronomico di Trieste, Via G. B. Tiepolo 11, I-34143 Trieste, Italy,
    \email{m.brazzini@inaf.it}
    \and
    Kavli Institute for Cosmology, University of Cambridge, Madingley Road, Cambridge, CB3 0HA, United Kingdom
    \and
    Cavendish Laboratory - Department of Physics, University of Cambridge, 19 JJ Thomson Avenue, Cambridge, CB3 0HE, United Kingdom
    \and 
    Department of Physics and Astronomy, University College London, Gower Street, London WC1E 6BT, United Kingdom
    \and 
    Steward Observatory, University of Arizona, 933 N. Cherry Avenue, Tucson, AZ 85721, USA
    \and 
    Max-Planck-Institut f\"ur extraterrestrische Physik, Gie{\ss}enbachstra{\ss}e 1, 85748 Garching, Germany
    \and
    Dipartimento di Fisica e Astronomia ``G.\  Galilei,'' Universit\`{a} di Padova, Vicolo dell'Osservatorio 3, I-35122 Padova, Italy
    \and
    INAF-Osservatorio Astronomico di Padova, Vicolo dell'Osservatorio 5 I-35122, Padova, Italy
    \and
    Jeremiah Horrocks Institute, University of Central Lancashire, Preston, PR1 2HE, UK
    \and
    Centro de Astrobiologia (CAB), CSIC-INTA, Ctra. de Ajalvir km 4, Torrejon de Ardoz, E-28850, Madrid, Spain
    \and
     IFPU — Institute for Fundamental Physics of the Universe, via Beirut 2, I-34151 Trieste, Italy
     \and 
     Scuola Normale Superiore, Piazza dei Cavalieri 7, I-56126 Pisa, Italy
     \and
     INAF – Osservatorio Astrofisico di Arcetri, Largo Enrico Fermi 5, I-50125 Firenze, Italy
     \and
     Dipartimento di Fisica e Astronomia, Università degli Studi di Firenze, via G. Sansone 1, 50019 Sesto Fiorentino, Firenze, Italy
     \and 
     Instituto de Astrof\' isica de Canarias, Calle V\'ia L\'actea, s/n, E-38205, La Laguna, Tenerife, Spain
     \and 
     Departamento de Astrofísica, Universidad de La Laguna, E-38206, La Laguna, Tenerife, Spain
     \and
     Space Telescope Science Institute, 3700 San Martin Drive, Baltimore, Maryland 21218, USA
     \and
     Department of Physics, University of Oxford, Denys Wilkinson Building, Keble Road, Oxford OX13RH, United Kingdom
}

% These dates will be filled out by the publisher
\date{Received XX January 2026; accepted XXX}

\abstract{
The population of Active Galactic Nuclei (AGN) newly discovered by the James Webb Space Telescope (JWST) exhibits peculiar properties that distinguish it from both local type I AGN and high-redshift quasars.
% , most notably an extreme X-ray weakness.
Most of these sources are compact, appearing as `little dots':
% in photometric surveys. 
among them, the sub-class (10-30\% of the total) characterized by significantly red optical colors has been named `Little Red Dots' (LRDs), while here we analogously introduce the term `Little Blue Dots' (LBDs) for 
%most of 
the remaining, bluer sources (70-90\%).
We then present a comparative analysis of the prototypical representatives  (`Rosetta Stones') of the two classes:
%of LRDs and LBDs respectively: 
GN-28074 at z=2.26, the Red Rosetta Stone, and GS-3073 at z=5.55, the Blue Rosetta Stone.

In both Rosetta Stones the broad Balmer lines are better described by exponential profiles rather than single Gaussians, similarly to normal low-redshift type I AGN, indicating that exponential profiles are not unique to LRDs.
They are both  extremely X-ray weak, show strong auroral \oiii$\lambda 4363$ emission, weak hot dust mid-IR emission, and no time variability. %indicating the presence of dust close to the sublimation temperature in the nuclear region.
However, they
%significantly
differ in terms of excitation diagnostics: the \heii$\lambda 4686$ line is undetected in the Red Rosetta but strongly detected in the Blue Rosetta in both narrow and broad components, with the latter much broader than hydrogen Balmer lines. This supports BLR stratification and disfavors the cocoon electron-scattering scenario.
An additional difference is the presence of prominent Balmer absorption in the Red Rosetta --indicative of extremely dense gas along the line of sight-- but absent in the Blue Rosetta.
%, as well as the detection of the \oiperm fluorescent emission in the Red Rosetta only.
%, but also in many normal type I AGN.  

Taken together, these results suggest that LRDs and LBDs share the same central engine as standard type I AGN, while differing in the amount and geometry of dense gas surrounding the accretion disk, and/or in their accretion properties.

}


\maketitle
%%%%%%%%%%%%%%%%%%%%%%%%%%%%%%%%%%%%%%%%%%%%%%%%%%

%%%%%%%%%%%%%%%%% BODY OF PAPER %%%%%%%%%%%%%%%%%%

\section{Introduction}

The James Webb Space Telescope \citep[JWST;][]{Gardner+} is revolutionizing our understanding of the formation and evolution of black holes in the early Universe. Indeed, thanks to its unprecedented sensitivity and wavelength range, JWST has opened a %totally
new discovery space by finding active galactic nuclei (AGN) with bolometric luminosities of $\log L_{\mathrm{bol}} [\mathrm{erg\,s^{-1}}]\sim 42 - 46$
\citep[e.g.,][]{Harikane+2023, Adamo+2025, Scholtz+2025_typeIIsample}, much lower than both low- and high-redshift quasars discovered with ground-based surveys. 
However, the new population of AGN discovered by JWST is not simply a scaled down version of luminous quasars, nor a simple extension to high redshift of the population of AGN found by previous surveys at later cosmic epochs.
Even the most securely identified broad line AGN are characterized by properties that are quite different from the populations investigated by previous studies, including: X-ray weakness \citep{Maiolino+2025, Yue+2024, Ananna+2024}, radio weakness \citep{Mazzolari+2024,Mazzolari+2025}, nebular emission lines whose excitation properties deviate from those observed in local AGN \citep{Harikane+2023, Ubler+2023, Maiolino+2024, Juodzbalis+2025, Zucchi+2025}, lack of the optical iron bump \citep{Trefoloni+2025}, and absent or seldom observed variability \citep{Kokubo+2024, Ji+2025a, Naidu+2025, Zhang+2025a}. 
These properties have motivated a large number of new models involving high gas obscuration \citep{InayoshiMaiolino2025,Juodzbalis+2024,Ji+2025a}, invoking scenarios characterized by high accretion rates \citep{Pacucci+2024,Madau+2024,King+2025}, and/or dense-gas cocoons fully enshrouding a rapidly accreting black hole, the `quasi-stars' \citep{Begelman2026} and `black-hole stars' \citep{Naidu+2025,deGraaff+2025}.

Within the population of JWST-discovered AGN, the class of so-called Little Red Dots (LRDs) has attracted most of the attention. 
%These are compact objects \citep[effective radii $\lesssim 100$~pc;][]{Furtak2023,Baggen+2024}
These sources are unresolved at rest-frame optical wavelengths, where a dominant point-like component often outshines the host galaxy,
and exhibit red optical colours and a characteristic v-shaped continuum \citep{Matthee+2024, Kocevski+2025}, pivoting at the Balmer limit \citep{Setton+2024} or close to it \citep{DeGraaff+2025b}. 
A large fraction of these photometrically selected LRD candidates display broad permitted emission lines \citetext{roughly 85\%, \citealp{Hviding+2025}; see also, \citealp{Greene+2024, Kocevski+2025}}, typical of type I, broad-line AGN. 
They often show Balmer absorption features \citep{Juodzbalis+2024,Matthee+2024, DEugenio+2025c,DEugenio+2025e,Deugenio+2025_irony,lin+2025b}, and also smooth Balmer breaks \citep{Setton+2024,Ji+2025a,Naidu+2025,deGraaff+2025, DeGraaff+2025b}, which have been interpreted as due to absorption by dense circumnuclear gas along the line of sight \citep{InayoshiMaiolino2025,Ji+2025a,Naidu+2025}.
Based on MIRI observations, LRDs show evidence for some hot dust, although the emission is generally weak \citep[e.g.,][]{Perez-Gonzalez+2024,Williams+2024,Casey+2025,Setton+2025, Delvecchio2025,Ronayne+2025}.

While remaining a compelling population, LRDs constitute less than 30\% of the AGN identified by JWST at redshift $z\sim5$ and at bolometric luminosities below the quasar regime \citep{Hainline+2025,Kocevski+2025,Taylor+2025}.
In contrast, the vast majority ($>$70\%) of these  spectroscopically confirmed, broad-line AGN exhibit blue optical and UV colours, i.e. both $\beta_\mathrm{opt}<0$ and $\beta_\mathrm{UV}<0$ \footnote{Where we define the slopes as $F_\lambda \propto \lambda^\beta$, and the measurements are made directly from the spectrum.}, similar to those of standard type I AGN and star-forming galaxies. 
These sources are predominantly compact \citep[e.g. ][who find that 60\% of JWST-discovered broad-line AGN are unresolved in NIRCam]{Hainline+2025,Juodzbalis+2025}, although we note that  -- just like in LRDs -- the presence of a more or less dominant unresolved component does not preclude the existence of an extended component, i.e. a host galaxy.

Despite being the primary population of JWST-discovered AGN, 
these blue counterparts have received less attention than LRDs. 
For instance, there are currently no studies that specifically investigate their UV--optical broad line spectral profiles and continuum emission, while these properties have been subject of intense scrutiny in LRDs \citep[e.g., ][]{Rusakov+2025, Naidu+2025, DEugenio+2025c, DEugenio+2025e, Deugenio+2025_irony, Chang+2025, Labbe+2024, Brazzini2025, Barro+2025}.
This is most likely due to the implicit assumption that these blue AGN
behave like normal AGN; however, they show remarkable peculiarities too, as discussed above, most notably their X-ray weakness. Therefore, their study is fundamental and complementary to the analysis of both standard luminous type I quasars and LRDs.

In this work, we present a comparative study of GN-28074 at $z=2.26$, the `Rosetta Stone' of LRDs \citep{Juodzbalis+2024}, and GS-3073, a blue, intermediate-luminosity AGN at $z=5.55$ \citep{Vanzella+2010, grazian+2020, Ubler+2023}, here adopted as the prototypical representative of the population of newly JWST-discovered blue AGN, which we denominate `Little Blue Dots' (LBDs) because of their blue colors and compact morphology (see next section for a more operational definition). 
% \RMcomm{Insert here our operational definition?}
% We stress that this is a preliminary definition; further insights into this newly identified AGN class, together with a more robust classification scheme, will be presented in a forthcoming paper (Geris et al., in prep.).
% archetypical `Little Blue Dot' (LBD).
%\mb{This choice is based on the assumption that LBDs correspond to JWST-discovered AGN with blue optical and UV slopes, compact morphologies, and X-ray and radio weakness. We stress that this is a preliminary definition; further insights into this newly identified AGN class, together with a more robust classification scheme, will be presented in a forthcoming paper (Geris et al., in prep.).}
% With this caveat in mind,
GN-28074 and GS-3073 serve as ideal archetypical cases, or `Rosetta Stones', of the LRD and LBD populations, respectively, enabling a detailed comparative analysis thanks to the high spectral signal-to-noise ratio (SNR) in their JWST spectra, and the excellent ancillary data warranted by their location in the GOODS deep extragalactic fields. 
% \citep{giavalisco+2005}.
%With this caveat in mind, both GN-28074 and GS-3073 are bright sources observed by JWST with high signal-to-noise ratio (S/N), exhibiting the full set of features characteristic of LBD and LRD populations, respectively. As such, they serve as prototypical cases, or `Rosetta Stones', that can be studied and compared in great detail.
%A more extensive statistical comparison using large samples of LBDs and LRDs (albeit with somewhat lower S/N) will be presented in a forthcoming paper. 

Throughout this work, we assume a flat $\Lambda$CDM cosmology with $\Omega_\text{m}=0.315$ and $H_0 = 67.4$~km/s/Mpc \citep{Plank2020}. 

\section{Little Red Dots and Little Blue Dots: definitions}
\label{section-definitions}

As in this paper we consider GN-28074 and GS-3073 as Rosetta Stones of Little Red Dots and Little Blue Dots, respectively, we shall first define these two classes.

\begin{figure}
    \centering
    \includegraphics[width=0.9\linewidth]{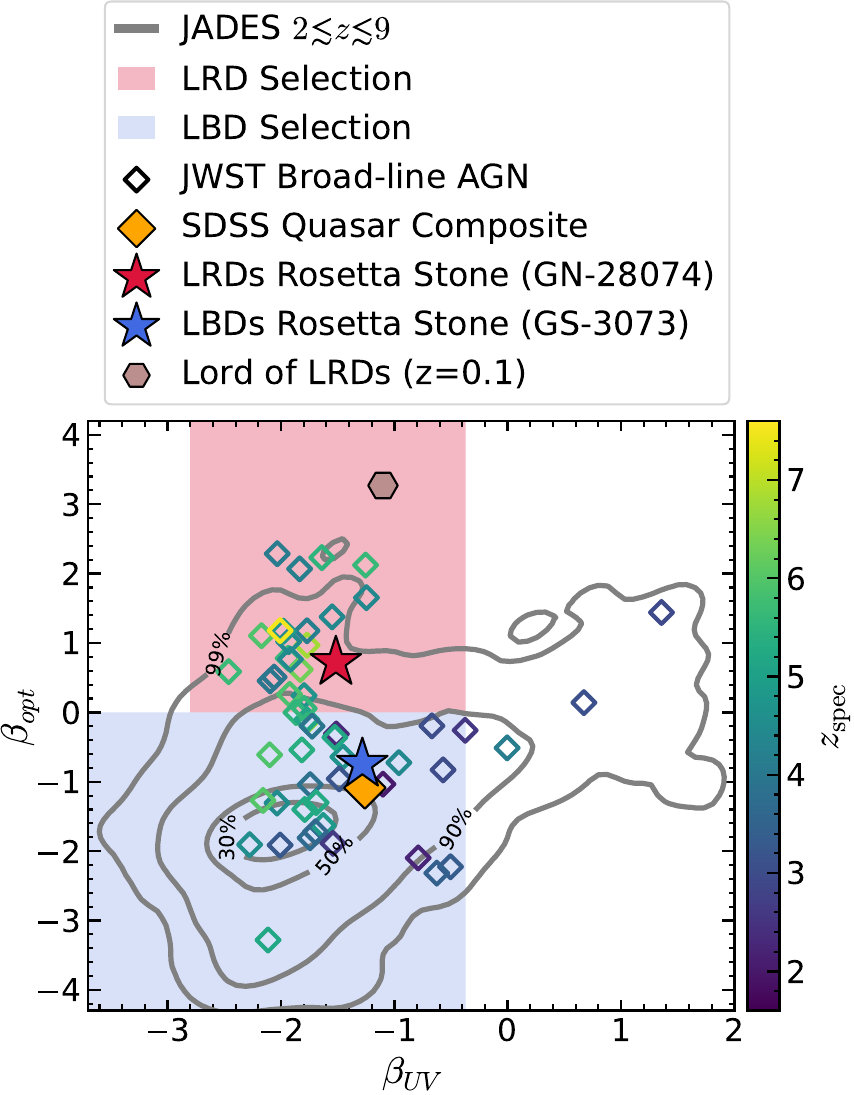}
    \caption{Rest frame optical versus UV continuum spectral slopes. While LRDs occupy a distinctive region in this plane (red shaded region), LBDs overlap
    with non-AGN galaxies at $z=2\text{--}0$ in the JADES survey \citep[grey contours;][]{Hainline+2025}, making their photometric
    selection at UV--optical wavelengths harder than for LRDs.
    The LRD selection, contours, and broad-line AGN (colour-coded by redshift) are from \cite{Hainline+2025}; the sample of broad-line AGN and the local
    `Lord of LRDs' are from \cite{Ji+2025_local,Kocevski+2024,Matthee+2024,Harikane+2023,Kokorev+2024,Sun+2025,Zhang+2025b,Juodzbalis+2025}.}
    \label{fig:continuum-slope-comparison}
\end{figure}

LRDs have been subject to extensive studies and there are various definitions. Many of them leverage the diagram shown in Fig.\ref{fig:continuum-slope-comparison}, which compares the slope of the optical continuum, $\beta _{opt}$, and the slope of the continuum in the UV, $\beta _{UV}$ \citep{Kocevski+2025,Hainline+2025}. In Fig.\ref{fig:continuum-slope-comparison} the contours show the distribution of all non-AGN galaxies in the JADES sample, mostly star forming galaxies, clustering around blue optical and UV slopes, but with a tail towards red UV and optical slopes, due to dust-reddened galaxies. The diamonds show all broad-line AGN in the JADES sample \citep[][Geris et al. in prep.]{Juodzbalis+2025}. The red shaded area shows the region with blue UV slopes and red optical slopes adopted by \citep{Kocevski+2025} to identify LRDs. Leveraging on this, we specifically follow a definition of LRDs following requirements \citep[e.g.][]{Kocevski+2025}:
\begin{itemize}
    \item Red rest-frame optical slope, $\beta _{opt}>0$ (Fig.\ref{fig:continuum-slope-comparison});
    \item Blue rest-frame UV slope, with $-2.8<\beta_{UV}<-0.37$
    (where the lower limit was adopted by \citealt{Kocevski+2025} to exclude brown dwarfs, while the upper limit is an attempt to avoid simply dust reddened systems);
    \item Compact (unresolved or barely resolved) morphology;
    \item Broad permitted emission lines (FWHM larger than about 1000~km/s) without a counterpart in the forbidden lines (such as \oiii), to exclude outflows, avoid contamination from compact post-starburst galaxies, and ensure the selection of AGN only.
    %To avoid contamination from compact post-starburst galaxies, and ensure that only AGN are selected, we also require broad permitted emission lines (FWHM larger than about 1,000~km/s), without a counterpart in the forbidden lines (such as [OIII]), therefore excluding an outflow.
\end{itemize}

Little Blue Dots have not been clearly defined in the past. They are, broadly, the vast majority of AGN selected by JWST at high-z and which are not LRDs, i.e. with both blue and optical slopes \citep[see also, e.g.,][]{Hainline+2025,Asada+2026}, with the exception of a few dust reddened AGN. However, being found by JWST is not a clear operational definition. As said, in terms of spectral optical/UV features they share properties similar to normal AGN (blue throughout the optical to UV spectral range). They are very compact, with their host galaxies barely detected; however, this is the case also for several luminous quasars. Yet, a remarkable difference with respect to normal AGN, is their extreme X-ray weakness \citep{Ubler+2023,Maiolino+2025}. We therefore adopt the following definition for Little Blue Dots:
\begin{itemize}
    \item Blue rest-frame optical slope, $\beta _{opt}<0$ (Fig.\ref{fig:continuum-slope-comparison});
    \item Blue rest-frame UV slope, with $\beta_{UV}<-0.37$
    (i.e. same upper limit as for the LRDs, to minimize selection of dust-reddened AGN, Fig.\ref{fig:continuum-slope-comparison})
    \item Compact (unresolved or barely resolved) morphology;
    \item Broad permitted emission lines (FWHM larger than about 1000~km/s) without a counterpart in the forbidden lines (such as [OIII]), to exclude outflows, hence identifying type I AGN
    \item X-ray weakness, with $k_\mathrm{bol} \coloneqq L_{\rm bol}/L_{2\text{--}10\,\rm keV}$ a factor of 10 higher than the $k_\mathrm{bol}- L_{\text{bol}}$ relation identified for ``normal'' type I AGN \citep{Duras2020,Maiolino+2025}.
\end{itemize}

In Fig.\ref{fig:continuum-slope-comparison} the blue shaded area illustrates the region of the diagram where LBDs are selected (together with the other requirements). This region largely overlaps with the distribution of star forming galaxies. Therefore, in contrast to LRDs, which can be easily and routinely selected through their colors, the selection of LBDs is much more difficult (despite them being more numerous than LRDs), as their colours overlap the plethora of star forming galaxies. Therefore, they can actually be identified only serendipitously via large spectroscopic surveys, through the detection of broad permitted lines. It is also interesting to note that normal, classical type I AGN mostly fall in this region - this is exemplified by the location of the SDSS Quasars Composite, marked with an orange-filled diamond \citep{VandenBerk2001}.

It should be noted that the broad line AGN identified in JADES (hollow diamonds) do not show a bimodal distribution, but spread continuously from the LBDs region to the LRDs region. This implies that probably LRDs and LBDs are not two separate populations, but a continuous distribution, whose classification in the two categories is somewhat arbitrary. This will be discussed more extensively in a separate paper (Geris et al. in prep.)

\section{Overview of `Blue' and `Red' Rosetta Stones}

Before providing our own analysis, we first clarify and justify the classification of GN-28074 and GS-3073 as representative of LRDs and LBDs, respectively, and summarize their properties based on previous studies.

\begin{figure}
    {\phantomsubcaption\label{fig:cutouts.a}
     \phantomsubcaption\label{fig:cutouts.b}}
    \centering
    \includegraphics[width=\columnwidth]{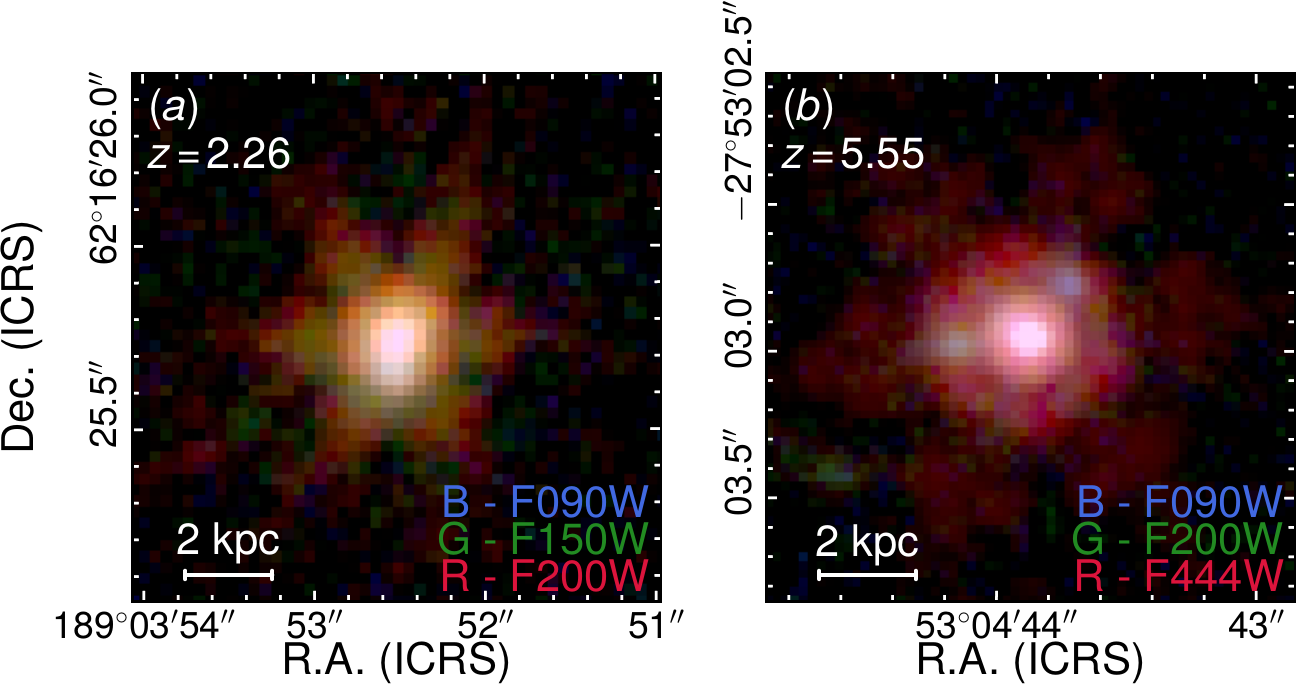}
    \caption{The `Rosetta Stone' of LRDs, GN-28074 \citep[panel~\subref{fig:cutouts.a}]{Juodzbalis+2024}, and GS-3073 (panel~\subref{fig:cutouts.b}), a luminous blue AGN \citep{Vanzella+2010,Ubler+2023}, taken here as the `Rosetta Stone' of `Little Blue Dots'. The false-colour RGB cutouts are from the JADES Collaboration \citep{Eisenstein+2023a,Rieke+2023,DEugenio+2025b}.}
    \label{fig:cutouts}
\end{figure}

\subsection{Spectroscopic data}\label{sec:data}

Spectroscopic observations of GN-28074 were carried out with the NIRSpec Micro-Shutter Assembly \citep[MSA;][]{Jakobsen+2022,Ferruit+2022}, as part of the JWST Advanced Extragalactic Survey (JADES) programme \citep[][programme ID $1181$]{Rieke+2020-JADES,Bunker+2020-JADES,Eisenstein+2023a,DEugenio+2025b,Curtis-Lake2025DR4,Scholtz2025RD4}, with data reduction described in detail in \cite{Juodzbalis+2024} and in the previous data-release papers \citep{DEugenio+2025b}. A detailed analysis of the spectral line profiles from the grating spectra has been presented in \citet{Brazzini2025}. Here we focus on the prism spectrum \citep[resolution $R=30\text{--}300$;][]{Jakobsen+2022}.

GS-3073 was observed with the NIRSpec Integral Field Unit \citep{Boker+2022} as part of the programme Galaxy Assembly with NIRSpec Integral Field Spectroscopy \citep[GA-NIFS, programme ID $1216$;][]{Boeker+2022-IFS}, and further details on data reduction and extraction  are provided by \cite{Perna2023_GANIFS} and \cite{Ubler+2023}. Throughout this work, we use both the prism and the G395H grating spectra, extracted from a 0.3" box centred on GS-3073, and already presented by \citet{Ubler+2023}. 
The nominal uncertainties on the aperture spectrum were upscaled by a factor of 2.5. As explained in \citet{Ubler+2023}, this ensures a good match between the standard deviation of the data (measured in featureless regions of the spectrum) and the noise vector. The most likely culprit for the mismatch in the nominal uncertainties is correlated noise in the spatial directions, which reduces the SNR of the aperture spectrum.

\subsection{GN-28074: The Rosetta Stone of Little Red Dots}

\begin{figure*}
    \centering
    \includegraphics[width=\linewidth]{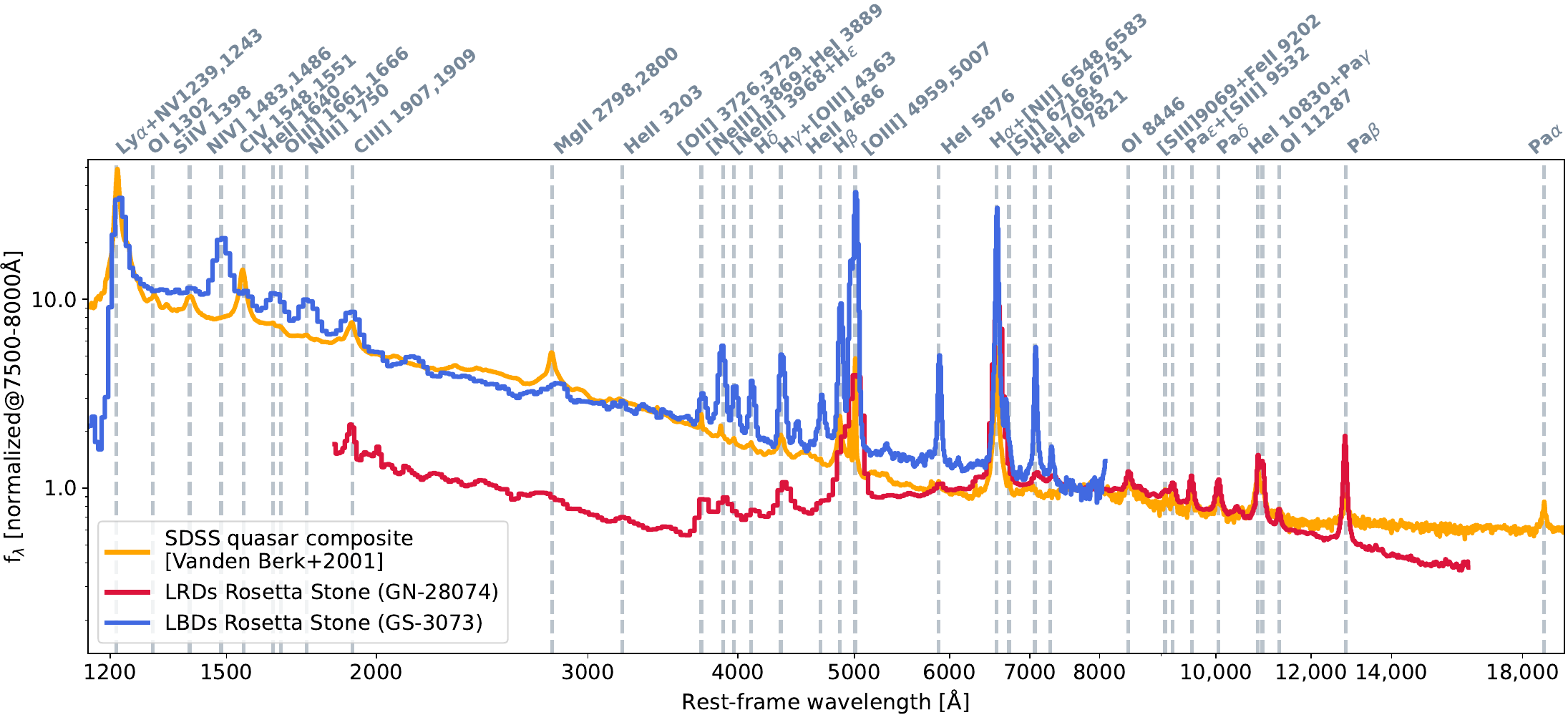}
    \caption{Spectral comparison between the Rosetta Stone of Little Red Dots \citep[GN-28074, from][red]{Juodzbalis+2024} and of Little Blue Dots \citep[LBDs; GS-3073, from][blue]{Ubler+2023}.
    LBDs share the blue UV--optical slopes of SDSS quasars \citep[yellow;][with the NIR extension from \citealp{Glikman+2006}]{VandenBerk2001}, but also show a number of
    differences (such as different emission-line ratios and EWs),
    meaning they are not scaled-down versions of quasars. The three spectra are normalized around $\lambda _{rest}\sim $8000~\AA\ (the longest wavelength in common to the three spectra). 
    }
    \label{fig:prism-comparison-spectra}
\end{figure*}

GN-28074 is a compact source in GOODS-North at redshift $z = 2.26$ (Fig.~\ref{fig:cutouts.a}), identified as
a broad-line AGN in \citet{Juodzbalis+2024}, and dubbed the `Rosetta Stone' of JWST-identified LRDs. % because it clearly displays all the characteristic features of this class of objects, as discussed above.
In addition to broad hydrogen lines, its LRD nature is confirmed by its v-shaped continuum, as seen in its prism spectrum, shown with a red curve in
Fig.~\ref{fig:prism-comparison-spectra} \citep[from ][]{Juodzbalis+2024}. 
Notably, the same figure highlights the remarkable difference between the spectrum of GN-28074 and those of both GS-3073 (see Section \ref{section-Gs3073-intro}), shown in blue, and the standard SDSS quasar composite, shown in yellow \citetext{from \citealp{VandenBerk2001}; note that we use the NIR extension from \citealp{Glikman+2006}}, which is adopted throughout this work as representative of standard type I AGN. 
More quantitatively, Fig.~\ref{fig:continuum-slope-comparison} shows the location of GN-28074 in the UV ($\beta_{\rm UV}$) versus optical ($\beta_{\rm opt}$) slopes diagram.
% , commonly used to identify LRDs \citep{Kocevski+2024,Kocevski+2025}.
% Spectroscopically confirmed broad-line AGN from JADES (empty diamonds) straddle most of the diagram, while the red shaded region marks the LRD selection criteria from \citet{Hainline+2025}. 
GN-28074 falls within the region adopted by many for selecting LRDs, with $\beta_{\rm UV, GN\text{-}28074} = -1.48\pm0.06$ and $\beta_{\rm opt, GN\text{-}28074} = 0.73\pm0.05$\footnote{We have used ad-hoc fitting windows: 2160--3100~\AA\ for the UV (masking \mgii), and 5360--6270~\AA\ for the optical (masking \hei 5876); this avoids strong emission lines and regions that are severely blended, due to the low spectral resolution of the NIRSpec prism.}.

Morphologically, \cite{Juodzbalis+2024} found that GN-28074 is compact in the rest-frame optical \citep[effective radius of $\sim 300$~pc;][]{vanderwel2014}, yet it is clearly spatially resolved in both the UV and \oiii emission, consistent with many LRDs, including spectroscopically confirmed ones at both low and high redshift \citep[e.g.,][]{Rinaldi+2025a, Torralba+2025a, Lin+2025_local,Ji+2025_local,DEugenio+2025e}.
It shows forbidden \feii emission \citep{Lin+2025_local,Ji+2025_local,Torralba+2025b,Deugenio+2025_irony} and \nai absorption \citep{Lin+2025_local,Ji+2025_local,Deugenio+2025_irony}, but no permitted \feiiperm or coronal lines such as \fevii.
Furthermore, it exhibits strong rest-frame
mid-infrared emission, which is detected even by Spitzer/MIPS at 24~\mum
\citep{Juodzbalis+2024}. 
This may appear uncharacteristic of many LRDs
\citep{Setton+2025,Perez-Gonzalez+2024,Williams+2024,Akins+2025,Casey+2025,deGraaff+2025}; however, more recent studies, especially at low redshift (hence offering more sensitivity to long wavelengths), have revealed that hot dust emission in LRDs is actually much more common than previously thought \citep{Lin+2025_local, Ji+2025_local, Delvecchio2025}.
In contrast to the MIR, GN-28074 remains undetected in X-rays \citep{Maiolino+2025}, with the 2-Ms Chandra data providing the most stringent upper limit to date on
$k_\mathrm{bol}$, about three orders of magnitude above the relation for normal AGN. 
We note that the Red Rosetta Stone would remain an outlier in the
$k_\mathrm{bol}\text{--}L_{\rm bol}$ diagram even if $L_\mathrm{bol}$ was overestimated by a factor of 100.

GN-28074 displays strong $n=2$ hydrogen absorption
of non-stellar origin \citep{Juodzbalis+2024}, a strong and smooth Balmer break \citep{Ji+2025_local}, and broad Balmer lines with non-Gaussian wings \citep{Juodzbalis+2024, Brazzini2025}, a feature common to several LRDs \citep{Rusakov+2025}, which has been interpreted as evidence for Doppler broadening by electron scattering \citep{Chang+2025,
Torralba+2025b}. 
However, in the case of GN-28074, a detailed comparative analysis of the spectral shapes across three hydrogen lines (\hb, \ha, and Pa$\beta$) reveals different line widths \citep{Brazzini2025}, which is inconsistent with a simple scattering screen, where the optical depth is effectively wavelength independent. 
However, recent works based on radiative transfer simulations \citep{Chang+2025, sneppen2026} succeed in reproducing such line differences in terms of more complex models including also Balmer scattering, and/or bound-free absorption, and/or a combination of inflow and outflow.

\subsection{GS-3073 - The Rosetta Stone of Little Blue Dots}
\label{section-Gs3073-intro}

GS-3073 was first identified by \citet{Vanzella+2010} with ground-based observations. 
An AGN interpretation was later explored in detail by \citet{grazian+2020} and \citet{barchiesi+2023}, based on the detection of Ly$\alpha$, Ly$\beta$, \niv, \nv, and \ovi UV emission lines. 
%\niv$1486$, \nv$1240$, and \ovi$1035$
In particular, the presence of high-ionization nitrogen and oxygen lines (with ionization potentials of 97.9 eV and 138.1 eV for \nv and \ovi, respectively) favoured the presence of an AGN, although the non-detection in X-rays left open the possibility of a stellar interpretation.
The type I AGN nature of this object was unambiguously confirmed by rest-frame optical spectroscopy with NIRSpec-IFU \citep{Ubler+2023} from the GA-NIFS 
%GTO 
programme, showing prominent broad \ha, \hb and \heii lines, and also other high ionization lines such as \fevii, although very weak \citep[][]{Ji+2024}. 
GS-3073 also exhibits ionized outflows, traced by the presence of a weak broad component detected in \oiii. 
This component, while broader than the narrow emission, is not sufficiently broad to be confused with BLR emission -- consistent with our definition of LBDs in Section \ref{section-definitions} -- and is present in all other emission lines \citep[Sect. \ref{section-emission-line-profiles-gs3073}; see also][]{Ubler+2023, Venturi+2025}.
Additionally, based on the extremely strong \niv and \niii seen in the prism spectrum, \citet{Ji+2024} determined a very high N/O abundance ratio in the high-ionization, inner regions \citep[see also][]{Ji+2025b}.
Indeed, high N/O may be a general feature of low-luminosity, high-redshift AGN, since spectral stacking of JWST-detected broad-line AGN reveals unusually bright \niii, which is not seen in (much deeper) stacks of non-AGN galaxies \citep{Isobe+2025}.
In contrast, among luminous AGN and quasars, the fraction of  `nitrogen loud' sources is
very small \citep[$\sim 1$~percent;][]{Baldwin+2003,Bentz+Osmer2004,Jiang+2008}.
Enhanced N/O could also explain the detection of \niv in the luminous LRD UNCOVER-45924 \citep{Labbe+2024,Torralba+2025a,Ji+2025b}, and the high \nv/\heii $\lambda1640$ ratio observed in two broad-line AGN at $z>7$ -- incidentally a LBD and a LRD \citep{Tang+2025}.

Morphologically, GS-3073 appears compact in the rest-frame optical continuum (Fig.~\ref{fig:cutouts.b}), and shows evidence for faint UV-blue neighbours -- not unlike LRDs \citep{Juodzbalis+2024,Killi+2024,Baggen+2024,Chen+2025,Rinaldi+2025a,DEugenio+2025e}.
However, GS-3073 is detected in the FIR,
%\citep[unlike LRDs, which remain undetected even in stacks;][]{Akins+2025,Casey+2025}, 
with
the ALPINE survey\footnote{Source ID: CANDELS\_GOODSS\_14} reporting \textup{[C\,\textsc{ii}]} 158$\mu$m emission, albeit the inferred molecular-to-stellar mass ratio
of 0.1 is among the lowest detected at this redshift \citep{barchiesi+2023}. 
This represents a significant difference with respect to LRDs, which remain undetected at these wavelengths, even in stacks \citep{Akins+2025, Casey+2025}. 
To date, only a single \textup{[C\,\textsc{ii}]} 158$\mu$m detection has been reported for a LRDs  \citep{Golubchik2025}, but the presence of a nearby companion complicates a definitive association of the emission with the LRD itself.

Fig.~\ref{fig:prism-comparison-spectra} shows the prism spectrum of GS-3073 (in blue), highlighting its blue spectral slope from the rest-frame UV to the rest-frame optical, similar to the SDSS quasar composite spectrum (in yellow), although the latter is slightly redder. 
Despite the comparable continuum slopes, the two spectra clearly exhibit striking differences, not only in terms of emission line intensities and widths, but also in terms of relative line ratios. This will be discussed in detail in a forthcoming paper. 
The similarity of optical and UV slopes of the Blue Rosetta GS-3073 ($\beta_{\rm UV, \rm GS-3073} = -1.30\pm0.01$ and $\beta_{\rm opt, \rm GS-3073} = -0.76\pm0.02$) clearly makes GS-3073 fall in the LBDs classification region of Fig. \ref{fig:continuum-slope-comparison}, and are also close to the bulk of type I AGN and star-forming galaxies -- which largely overlap. 
% As a consequence, LBDs are typically identified serendipitously through spectroscopy rather than colour selection, in contrast to the now routine photometric selection of LRDs.
It is interesting that in terms of $\beta_{\rm UV}$, GS-3073 is as blue as the SDSS type I composite ($\beta_{\rm UV, SDSS} = -1.240\pm0.006$ and $\beta_{\rm opt, SDSS} = -1.08\pm0.01$).

Lastly, GS-3073 is extremely X-ray weak, remaining undetected in the 7-Ms Chandra campaign in GOODS-South \citep{Luo+2017}. Due to its luminosity, the constraints on $k_\mathrm{bol}$ are also very strong, resulting into one of the AGN with the highest lower limit on $k_\mathrm{bol}$ in the whole sample explored by \cite{Maiolino+2025}.

\section{Time variability}\label{sec:timevar}

Multiple studies have demonstrated a lack of LRD variability on relatively short timescales (months to years; \citealt{Kokubo+2024, Ji+2025a, Naidu+2025, Zhang+2025a,Burke+2025}). We explore the year-long variability of GN-28074 by comparing JWST NIRCam F090W, F115W, and F356W photometry measured from the JADES 1181 data and the Complete NIRCam Grism Redshift Survey data (CONGRESS, \citealt{Egami+2023}, programme ID 3577). The JADES data (epoch1) was obtained on UT 3 February 2023, and the CONGRESS data (epoch2) was obtained on UT 13--18 February 2024. For each filter and epoch combination, we measure source flux with an 0.1'' aperture and estimate flux uncertainty with 500 randomly placed 0.1'' apertures around GN-28074. In each of the three comparison filters, the epoch1 and epoch2 flux measurements agree within 1$\sigma$. Therefore, similar to other LRDs, GN-28074 does not exhibit variability on year-long timescales.

While there was no multi-epoch JWST/NIRCam coverage of GS-3073, we compared NIRCam F150W photometry measured from the Observing All Phases of Stochastic Star Formation (OASIS, \citealt{Looser2024}, programme ID 5997) program to HST/WFC3 F160W photometry measured from the Hubble Legacy Fields (HLF; \citealt{Whitaker2019}). The OASIS imaging was obtained on UT 15--19 October 2024, and the HLF GOODS-S image is a composite image containing many epochs of HST imaging taken prior to 2020. We adopted photometry from the JADES photometric catalogs in the public Data Release 5 \citep{Robertson+2026,Johnson+2026}. The photometry was measured with a 0.1'' aperture, and photometric uncertainty was estimated with randomly-placed apertures in source-free regions. The two fluxes agree at the 3$\sigma$ level, indicating no variability in the few years between the HLF images and OASIS image. The reason why the fluxes do not agree at the 1$\sigma$ level may be attributed to the slight difference in comparison filters.

\section{Mid-infrared emission}\label{section-mid-IR}

\begin{figure}
    {\phantomsubcaption\label{fig:mir.a}
     \phantomsubcaption\label{fig:mir.b}}
    \centering
    \includegraphics[width=\linewidth]{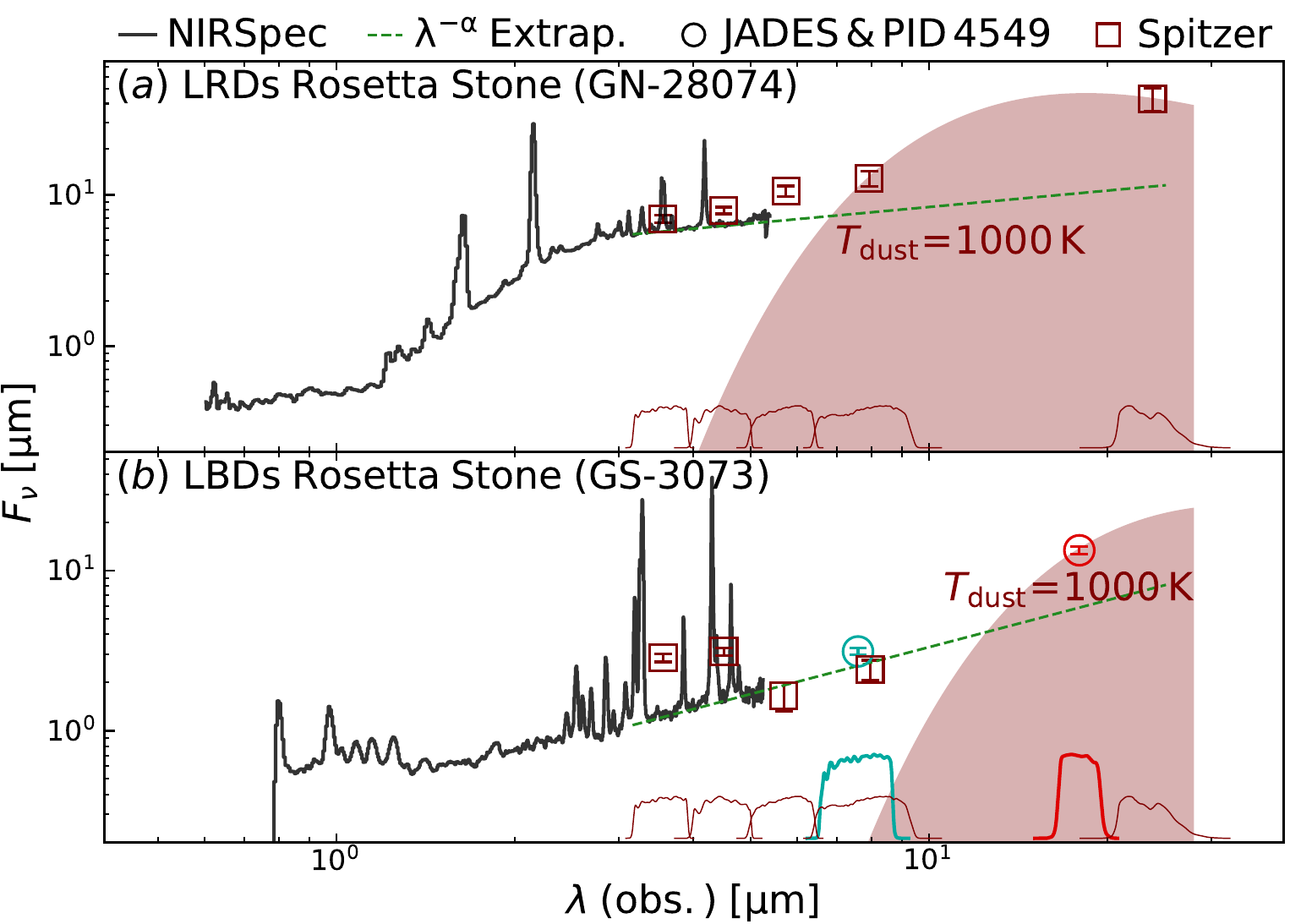}
    \caption{Comparison of rest-frame MIR emission between the Rosetta Stone of LRDs (panel~\subref{fig:mir.a}) and of LBDs (panel~\subref{fig:mir.b}). The solid black
    curves are from NIRSpec/prism, while the green line is a power-law extrapolation of the emission-line subtracted continuum. The observed MIR photometry clearly lies above the extrapolation, and can be explained by thermal emission from dust at
    $T_\mathrm{dust}=1,000$~K. The data points (and matching filter transmission curves) are Spitzer IRAC and MIPS24 (brown), JADES MIRI F770W (blue),
    and GO-4549 MIRI F1800W (red).}
    \label{fig:mir}
\end{figure}

To explore hot-dust emission, we complement NIRSpec with archival photometry from Spitzer, collected from the
Rainbow Database, and with JWST/MIRI photometry from JADES \citep[JWST Advanced Deep Extragalactic Survey;][]{Eisenstein+2023a} and PID~4549 (PI G.~Rieke; priv. comm.). The data are shown in Fig.~\ref{fig:mir}, where we compare the NIRSpec data (black), a power-law 
extrapolation of NIRSpec (green dashed), and the MIR data.

The filled curves are blackbody flux density models, with temperatures taken from \citet{Lin+2025_local}, and scaled to match the spectrum and/or photometry.
It is clear that both GN-28074 and GS-3073 require a hot-dust component, similar to
what is seen in normal AGN and quasars (although we are unable to constrain the dust temperature in
GS-3073).
It has been reported that the hot dust in LRDs is deficient compared to the expectations from a power-law fit to the optical plus dust attenuation \citep{Setton+2025}. Our comparison shows that both LRDs and LBDs can
in fact contain hot dust. This result is in line with the findings of \cite{Lin+2025_local} and \cite{Ji+2025_local}, who report significant hot dust emission in local LRDs. This suggests that previous claims of an absence of hot dust may instead derive from limited sensitivity of MIRI at high redshift, which could have prevented the detection of this component. This issue is exacerbated by the fact that, at high redshift, most MIRI bands probe only the Wien side %the Rayleigh–Jeans tail
of the blackbody SED, making the inferred emission highly sensitive to the assumed dust temperature and emissivity.
More importantly, Fig.~\ref{fig:mir} illustrates that at z$>$4 it is extremely difficult to assess the presence of hot dust without sensitive data at $\lambda_{\textrm{obs}}>4\mu$m.

In line with our findings, the recent, extensive and detailed study by \cite{Delvecchio2025} indeed find that a large fraction of LRDs show clear evidence for hot dust emission.

Our findings suggest that: 1) dust is actually present in the nuclear region of both LRDs and LBDs; and 2) it is hot ($\sim 1000~K$) as typically observed in AGN, meaning that such dust must be exposed to the UV ionizing source and close to the sublimation temperature. This implies that certainly the UV radiation produced by the AGN must be somehow escaping the central region, in contrast with the black hole star simple scenario.

In the Discussion we investigate more quantitatively the prominence of hot dust emission in Red Rosetta and Blue Rosetta, and the possible implications.

\begin{figure*}[th]
\centering
    \begin{subfigure}{0.95\linewidth}
        \includegraphics[width=\linewidth]{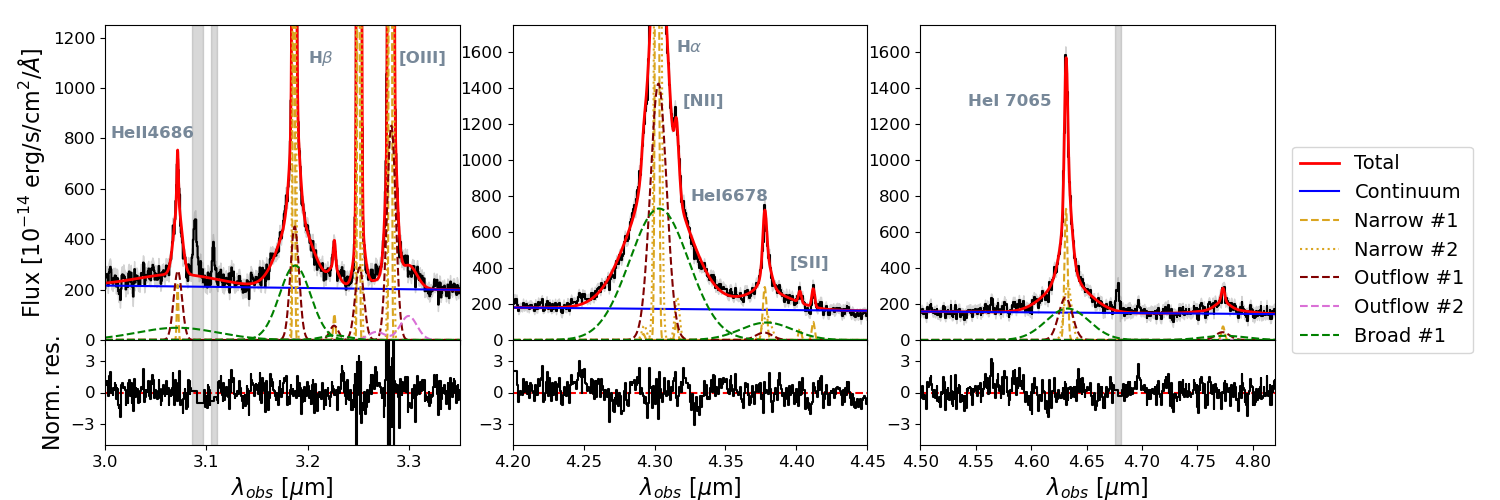}
        \caption{Single broad Gaussian model.}
        \label{fig:gs3073-singlebroadgauss}
    \end{subfigure}
    \vspace{-0.1cm}
    \begin{subfigure}{0.95\linewidth}
        \includegraphics[width=\linewidth]{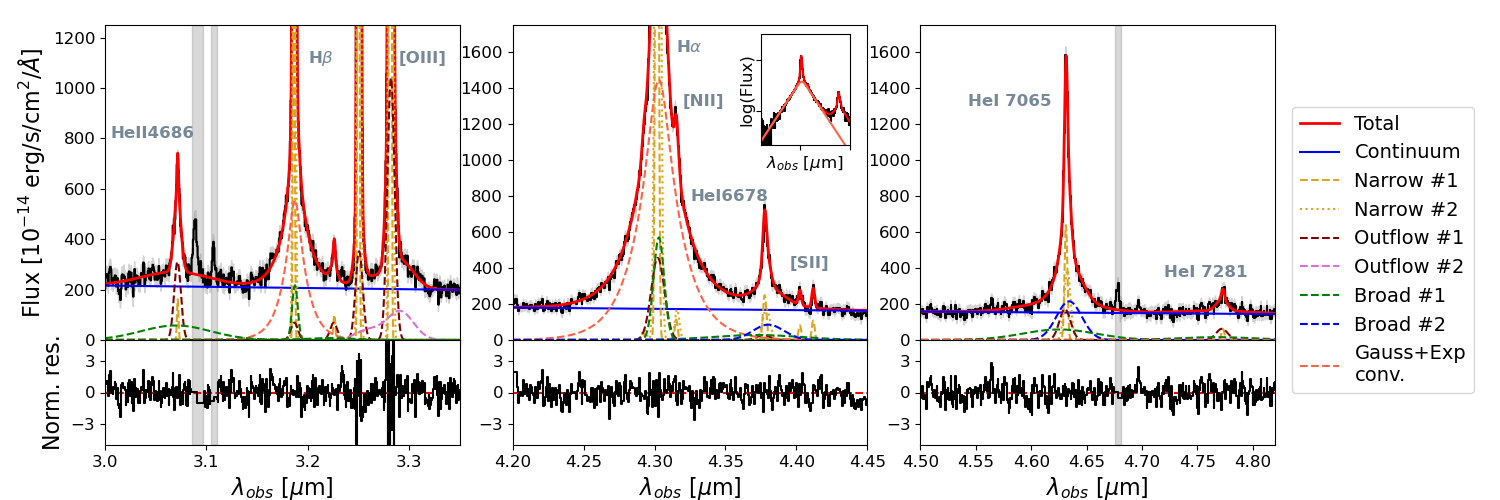}
        \caption{Exponential model.}
        \label{fig:gs3073-expmodel}
    \end{subfigure}
    \vspace{-0.1cm}
    \begin{subfigure}{0.95\linewidth}
        \includegraphics[width=\linewidth]{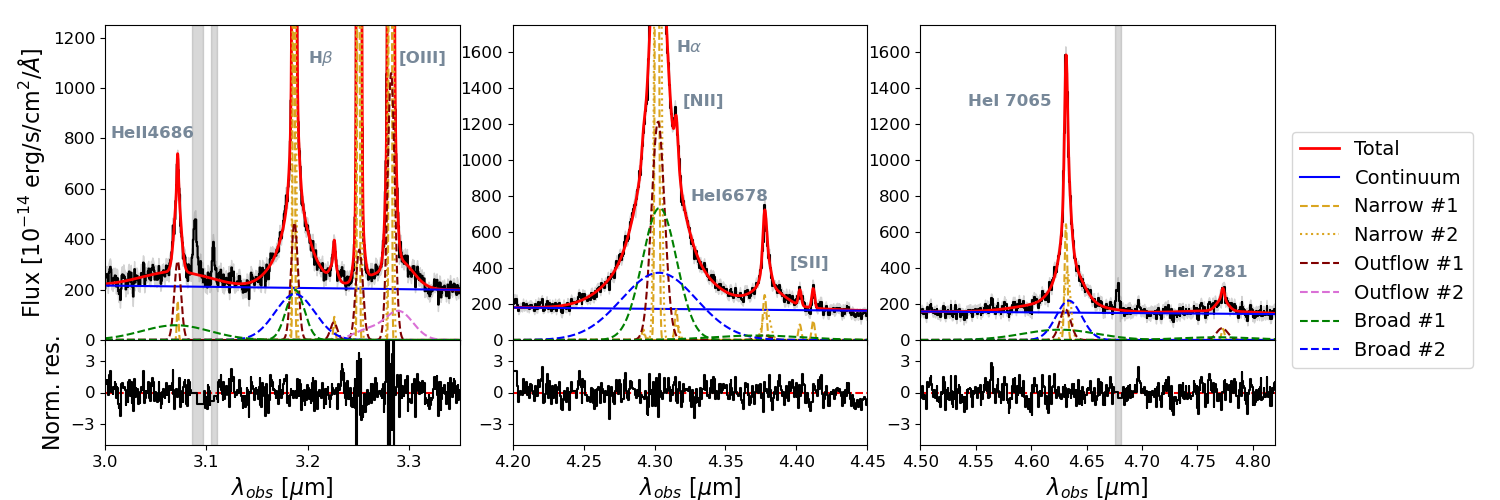}
        \caption{Double broad Gaussian model.}
        \label{fig:gs3073-doublebroadgauss}
    \end{subfigure}
\caption{Spectrum of GS-3073, showing the relevant hydrogen, helium and metal emission lines analysed in this work. 
The total best fit is reported in solid red for each of the three different broad emission models considered in this work (from top to bottom): single Gaussian, exponential, double Gaussian. Individual line components are reported in dashed colours, and the underlying continuum in solid blue. 
For each model, the lower panels display the residuals between the bestfit model and the observed spectrum, normalised by the spectral uncertainties. The masked regions (in gray) correspond (from left to right) to the blend of \hei$\lambda 4711$ and \ariv$\lambda 4713$, \ariv$\lambda 4740$ and \ariii$\lambda 7136$ lines.}
\label{fig:gs3073-linefits}
\end{figure*}

\begin{figure*}
\centering
    \begin{subfigure}{0.32\linewidth}\includegraphics[width=\linewidth]{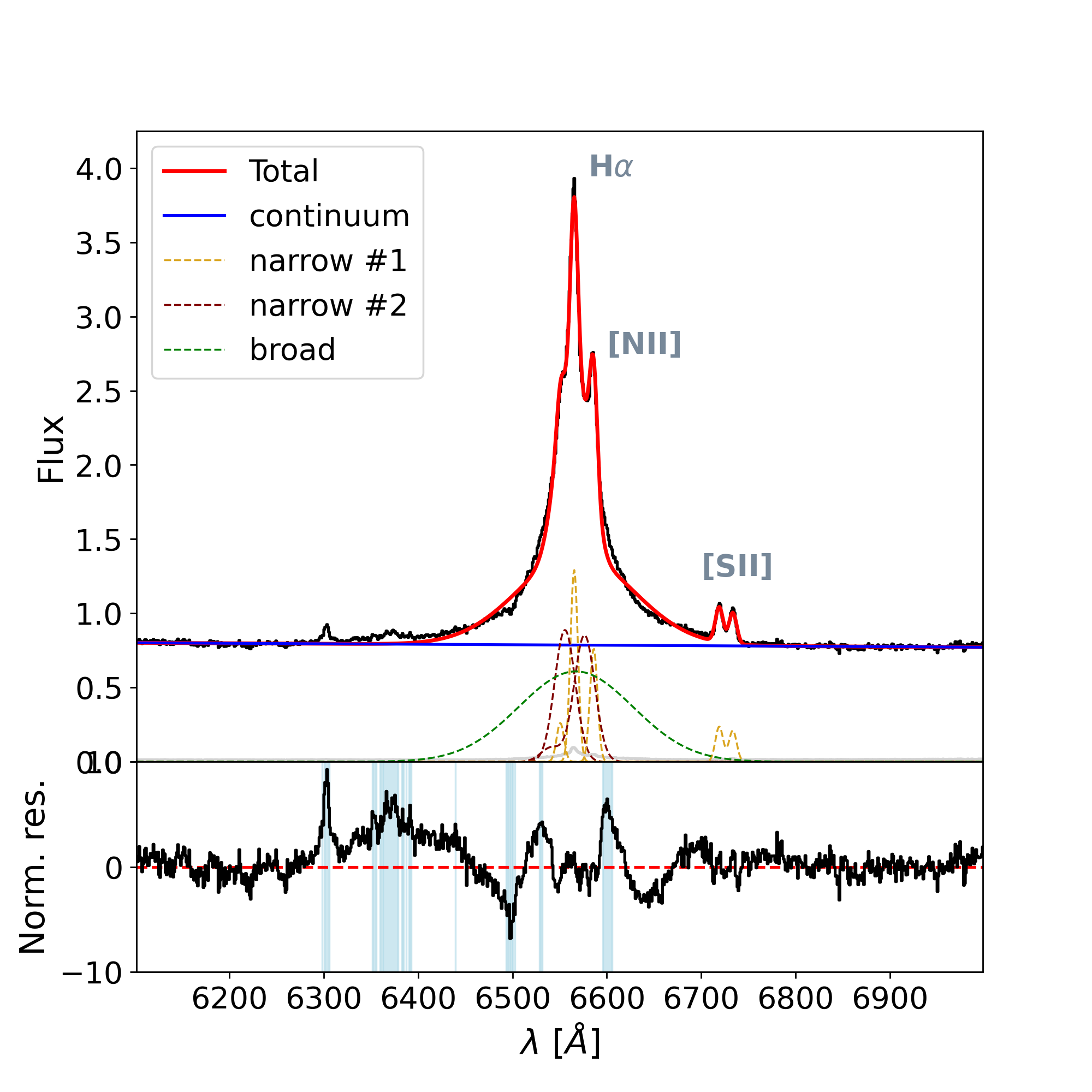}
    \end{subfigure}
    \begin{subfigure}{0.32\linewidth}
        \includegraphics[width=\linewidth]{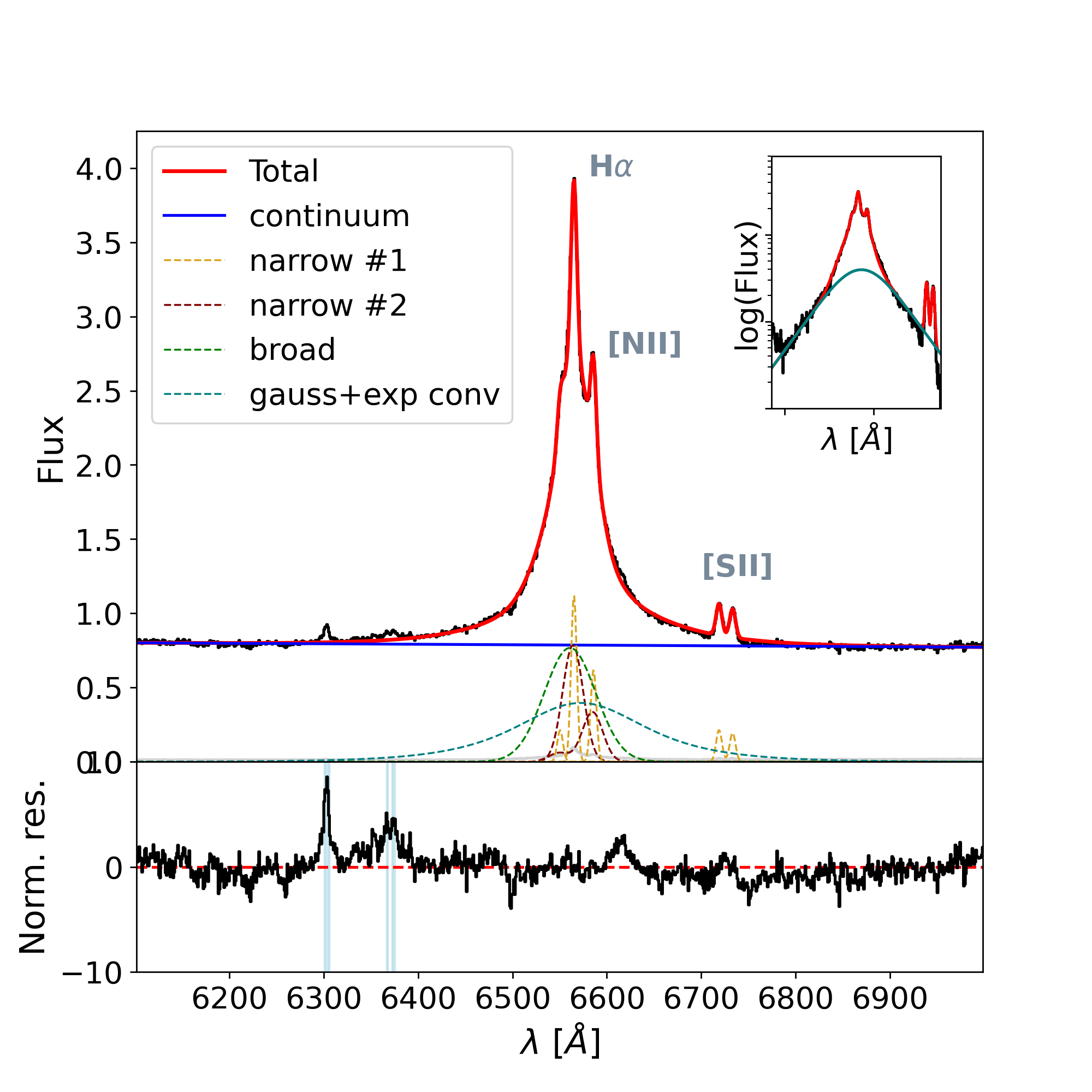}
    \end{subfigure}
    \begin{subfigure}{0.32\linewidth}
        \includegraphics[width=\linewidth]{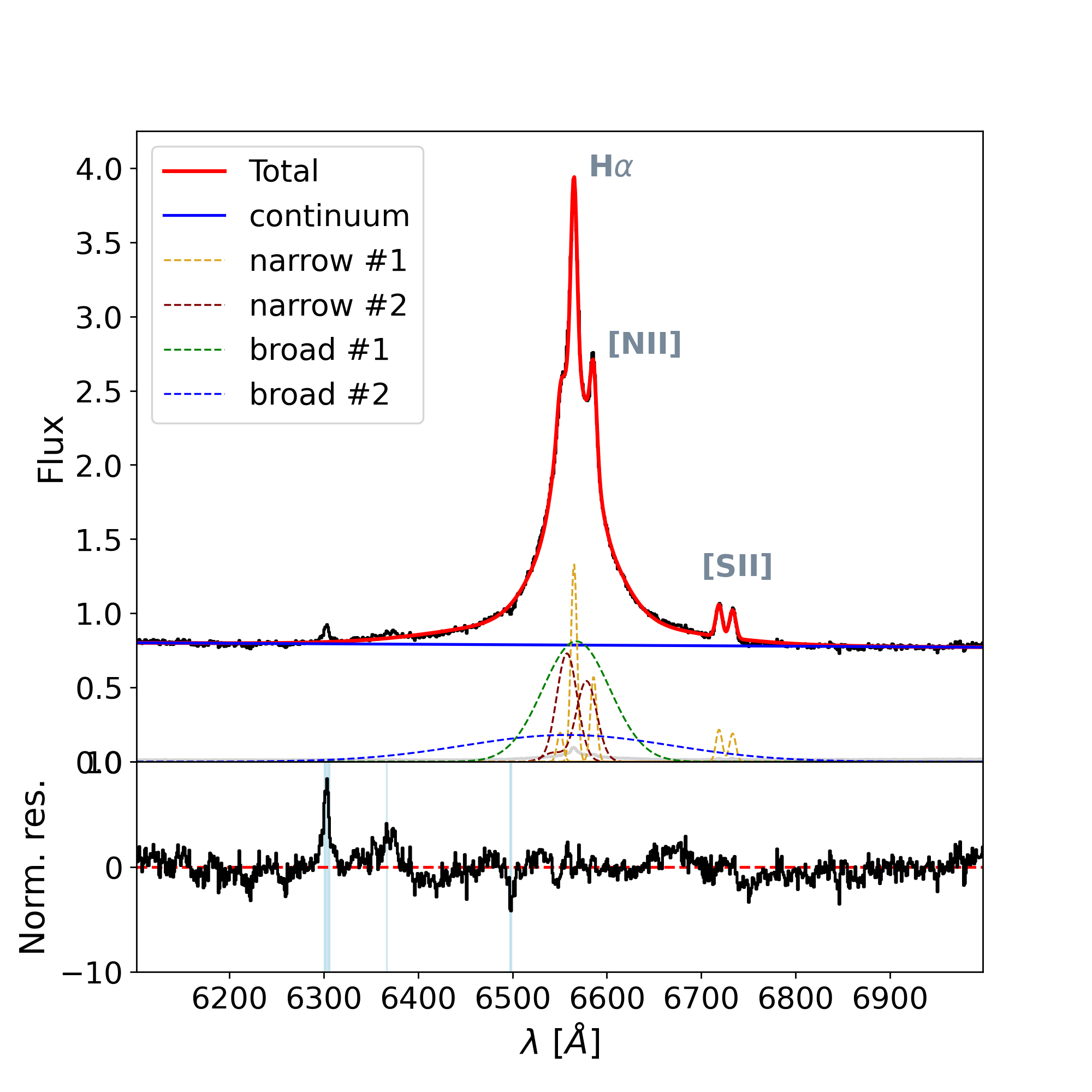}
    \end{subfigure}
\caption{Fit of the \ha complex in the composite spectrum from \cite{VandenBerk2001} using the three different models for broad emission analysed in this work (see Sect. \ref{section-emission-line-profiles-gs3073}), from
left to right: single Gaussian, exponential, double Gaussian. 
In the lower panels, data points differing from the model more than $4 \sigma$ are highlighted in blue. 
Some of these points are associated to emission lines not considered in our model, for instance the doublet \oi$\lambda\lambda6300, 6363$ is clearly detected in the residuals of all the considered models, and some weak emission from \oi or \fex at $\sim 6370$~\AA, as also reported in \cite{VandenBerk2001}. }
\label{fig:vandenberk+2001_composite_template_fits}
\end{figure*}

\section{Emission line profiles in the Blue Rosetta (GS-3073)}
\label{section-emission-line-profiles-gs3073}

In this section, we analyse the emission-line profiles in GS-3073 only -- the Blue Rosetta --, and we refer to \citet{Brazzini2025} for the analysis of GN-28074 -- the Red Rosetta -- using the same methods as here.
We focus on permitted lines observed in the high-resolution G395H spectrum, i.e. hydrogen and helium lines: Balmer (\ha, \hb), singly ionized He (\hei$\lambda 4922,\lambda6678,\lambda7065,\lambda7281$) and doubly ionized He (\heii$\lambda 4686$). Metal lines (e.g., \nii$\lambda \lambda 6548,6583$, \sii$\lambda \lambda 6716,6731$) are considered only insofar as they are necessary to accurately model nearby lines of interest.
We refer to \cite{Ubler+2023} and \cite{Ji+2024} for a comprehensive list of emission line detections, including those not analysed in this work.
Throughout the analysis we use the Bayesian Bic Criterion \citep[BIC;][]{Schwarz1978,liddle2007} with a threshold of $\Delta \textrm{BIC}=10$ to assess the statistical preference of different models. 

Regarding \hei emission, we mask \hei$\lambda 4711$ due to its weakness and strong blend with \ariv$\lambda 4713$, and we exclude \hei$\lambda5876$ from our analysis because of the presence of a continuum feature that is difficult to model. This feature skews the broad-Gaussian component, pushing for an extremely broad ($\sim 15,000 $ km/s) and blue-shifted (by almost $2000$ km/s) solution which is not seen in any other broad line. Since this feature is seen only in the G395H data but not in the prism, it may arise from incorrect background subtraction in this spectral region, for instance due to a bright spoiler source or to a `disobedient' shutter in the NIRSpec/MSA \citep{Jakobsen+2022,Ferruit+2022}.

For each emission line, we model the local continuum as a linear function and the line emission as a superposition of Gaussian functions: one outflow component, as seen in \oiii (see discussion below); two narrow components; and two broad components, adopted after verifying that a single broad component does not provide an adequate fit ($\Delta \textrm{BIC} = \textrm{BIC}_{\textrm{single\_broad}} -\textrm{BIC}_{\textrm{double\_broad}} =  125$,  Table \ref{tab:statistics}; see also top panel of Fig. \ref{fig:gs3073-linefits}), at least for the brighter lines (but see Section \ref{section-heii-profile} for a more detailed discussion on the broad profile of \heii$\lambda4686$ line). 
The need for a double narrow component in all lines --but the weak \heii$\lambda 4686$ and \sii doublet-- is dictated by the poor fit obtained using a single one ($\Delta \textrm{BIC} = \textrm{BIC}_{\textrm{single\_narrow}} -\textrm{BIC}_{\textrm{double\_narrow}} =  127$). 

We adopt a non-informative prior on the centroids of the various narrow components, therefore allowing for shifts by up to $\pm 500$ km/s.
We constrain the line full-width at half maximum (FWHM), respectively, between $50$~ and $700$~ km/s for the narrow components, and between $500$~ and $15,000$~ km/s for the broad and outflow components.
Given the similarity of \ha and \hb profiles, and likewise among the \hei profiles, we tie both the narrow and broad line widths within each group separately. Unlike GN-28074, GS-3073 has no evidence of Balmer absorption, which is therefore not included in our model.
The fitting results are shown in Fig. \ref{fig:gs3073-linefits}, lower panel, and line widths are reported in Table \ref{tab:line_widths_gs3073}.

Narrow components in all lines exhibit widths of a few hundreds of km/s and small relative velocity shifts of $\lesssim 50$~km/s. 
Two broader components are detected in \oiii, with FWHM$\sim 760$~ and $\sim 3060$~km/s, respectively.
The former is mildly blue-shifted with respect to narrow emission ($<100$~km/s from the bluest narrow component), while the latter is significantly redshifted (by $\sim 770$~km/s).
We interpret these components as likely signatures of outflows. 
Both of them were initially included in the fit of the other emission lines; however, the broader and more blue-shifted component contributed a negligible flux and was therefore excluded from the modelling of the remaining lines for simplicity. 
Its absence in other emission lines leaves open the possibility that this feature, rather than tracing an outflow, may correspond to faint, broad \feiiperm emission \citep{Trefoloni+2025}. 
In contrast, the narrower outflow component is clearly detected in the hydrogen and helium lines -- but not in metal lines. %, which we model with a single Gaussian component tied in kinematics to the broader of the two narrow Balmer components. 
We verified that the inclusion of such outflow component is required to obtain a satisfactory fit of hydrogen and helium lines, as removing it leads to a significant degradation of the fit quality, as established by studying both $\Delta \textrm{BIC} = \textrm{BIC}_\textrm{double\_broad,no\_outflow} - \textrm{BIC}_\textrm{double\_broad+outflow} = 269$ and the reduced chi-squared statistics ($ \chi^2_{\textrm{red}}=1.28$ versus $ \chi^2_{\textrm{red}}=1.05$ for our fiducial double Gaussian model including the outflow component).
%nb the relative bic variation would be deltaBIC/BIC=(2146-1877)/1877=14%, where deltaBIC = bic(double-broad-gauss-no-outflow - double-broad-gauss+outflow)
We note that, given its relatively narrow width and small velocity shift, this component that we dubbed as `outflow' does not conflict with the earlier definition of LBDs, which requires the absence of broad outflows with FWHM exceeding 1000~km/s. Indeed, it may alternatively be interpreted as an additional narrow component, possibly associated with satellites, that, combined with the other two, traces the complex gas kinematics within the galaxy.
However, since the focus of our work is not on narrow or outflow emission, we do not discuss any further the physical interpretation of these narrow components.
We just point out that two narrow components (at least) are necessary to achieve a satisfactory fit of hydrogen and helium lines, as the removal of one of them would significantly worsen the fit ($\Delta\textrm{BIC}>100$), and cause one of the broad components to collapse into a narrow one.
%deltaBIC/BIC=(1990-1877)/1877=6% and sigma_broad2 = 115 km/s
Such an intricate kinematic structure is plausible, given the presence of two likely companions identified in the photometry, as discussed above.

\subsection{Broad components of the hydrogen Balmer lines}
\label{section-exp-model}

The top panel of Fig.~\ref{fig:gs3073-linefits} shows that a single Gaussian fit to the broad Balmer lines leaves some significant residuals.
An effective model with two broad Gaussian components provides a substantially better fit (Fig. \ref{fig:gs3073-linefits}, bottom panel), with an improvement in the line statistics of  $\Delta \textrm{BIC} = \textrm{BIC}_{\textrm{single\_broad}}-\textrm{BIC}_{\textrm{double\_broad}} = 125$, i.e. almost 10\% (Table \ref{tab:statistics}). 
Both broad components are well constrained, with FWHM of $\sim 2000$~km/s and $\sim 4400$~km/s, respectively, and comparable fluxes. The total broad profile exhibits an effective width of $\sim 2500$~km/s. 

There are several studies claiming that broad line profiles of LRDs are characterized by exponential wings \citep{Rusakov+2025,Torralba+2025b,Nikopoulos+2025,sneppen2026}.
As we will discuss later, this feature does not really seem to be a peculiarity of LRDs; in this section, we explicitly explore this possibility.
To this end, we fit the broad component by adopting the same model used for GN-28074 in \cite{Brazzini2025}. 
Following their prescriptions, the broad Balmer emission is represented as the sum of a broad Gaussian component and the convolution of this broad Gaussian with a symmetric double exponential profile \citep{Rusakov+2025}, representing the unscattered and scattered BLR emission respectively. 
The parametrization of helium and other metal lines is left unchanged, i.e., pure Gaussians.
Given the similarity of the \ha and \hb profiles, we impose the same exponential width, $W$, and scattered fraction, $f_{\text{scatt}}$ (i.e. the fraction of BLR emission that undergoes at least one scatter within the dense gas envelope), in the two lines.
This assumption yields good results (middle panel of Fig. \ref{fig:gs3073-linefits}, Table \ref{tab:line_widths_gs3073}), with $W \sim 1000$~ km/s, consistent with the typical values found in LRDs \citep{Rusakov+2025, Brazzini2025}, $f_{\text{scatt}}= 0.91$, and FWHM$_\textrm{broad}\sim 600$~km/s. 
The exponential profile is highly preferred relative to the single Gaussian, with a $\chi^2_\textrm{red}=1.05$ for the exponential profile vs $\chi^2_\textrm{red}=1.18$ for the single Gaussian, and with a $\Delta \textrm{BIC} = \textrm{BIC}_{\textrm{single\_broad}} - \textrm{BIC}_{\textrm{exp}} = 139$ (Table \ref{tab:statistics}).
 Instead, the statistical difference between double Gaussian and exponential fits is marginal: they display the same $\chi^2$ and $\chi^2_{\textrm{red}}$ values, and the relative BIC variation is less than 1\%, with the exponential model slightly preferred. 
 %We find that, although for the single \ha line the exponential model is statistically favoured, when considering multiple Balmer lines the double Gaussian model is slightly preferred, with $\Delta$BIC=68, which decreases to 60 when masking all outliers at $> 4 \sigma$ (blue-shaded regions in the residuals panels of Fig. \ref{fig:gs3073-linefits}).
This implies that these two models can hardly be differentiated using any statistical metrics, because at these extremely low-level effects such as minor flat fielding, background subtraction and calibration effects probably dominate the residuals. 
This is a general warning when attempting to differentiate between exponential, double/multiple-Gaussian, but also Lorentzian and power-law profiles.

The adopted model for the broad Balmer emission has no significant impact on the helium or metal lines, and the widths of the narrow Balmer components remain consistent within $1 \sigma$.

\subsection{Broad components of the helium lines}
\label{section-heii-profile}

Similar to the Balmer lines, we model helium lines with one (\heii) or two (\hei) narrow, one outflow, and one (\heii) or two (\hei) broad components. 
For \hei, the kinematics of the narrow components are left free. 
For \heii, the kinematics of the single narrow component are tied to that of the narrower Balmer lines (FWHM of $\sim 150$~ km/s). 
In both \hei and \heii lines, the outflow is tied to \oiii, while the parameters of the broad component(s) are left unconstrained in order to explore similarities and
differences with respect to lower-ionization lines.

The decision to model the narrow and broad \heii emission with single Gaussian components is motivated by the faintness of the line: this simpler parametrization reduces parameter uncertainties and degeneracies without significantly affecting the fit quality ($\Delta \textrm{BIC} = \textrm{BIC}_{\textrm{\heii\_double\_broad}} - \textrm{BIC}_{\textrm{\heii\_single\_broad}} = 22$, implying a relative improvement of just $\sim 1\%$).
In contrast, \hei broad lines show a clear preference for a profile more complex than a single Gaussian, with exponential or double Gaussian performing equally well.

While \hei lines exhibit broad profiles comparable in width to the Balmer lines, the broad \heii4686 emission exhibits a total width of $\sim 7000$~km/s, driven by the presence of an extremely broad (FWHM$\sim8000$~km/s), slightly blue-shifted (by $\sim 200$~ km/s) component.
This is consistent with the BLR ionization stratification seen in reverberation studies \citep{Bentz2023, McDougall2025, xi2025}, and will be explored more in detail in the next sections.
% to address Bartolomeo's comment, we could add this text, but perhaps it's better in the discussion than here?
% where the `optimally emitting' \heii clouds are expected to be closer to the SMBH due to the higher ionization potential of \heii

\section{Discussion}

In this section, we compare the properties of GN-28074 and GS-3073, highlighting their commonalities and differences.

\subsection{Hydrogen broad-line emission}

Both GN-28074 and GS-3073 exhibit broad Balmer lines (FWHM of few thousands km/s, depending on the adopted emission line model, Table \ref{tab:line_widths_gs3073}).
The high luminosity 
%density 
of these lines implies powering by an accreting SMBH \citep[e.g.,][]{Maiolino+2024}, classifying both LRDs and LBDs as specific sub-classes of broad-line AGN.
Furthermore, in both sources the spectral profiles of the broad hydrogen lines are markedly non-Gaussian, especially \ha. 
%and if considering \ha alone, an exponential model is statistically always preferred. 
This is crucial, because most current LRD observations cover only broad \ha, since \hb is much harder to detect due to the steep Balmer decrement, with \ha/\hb$>3$, typically of order 10  \citep[e.g.,][]{Juodzbalis+2024,DEugenio+2025c,Deugenio+2025_irony,Torralba+2025b,Labbe+2024}, and up to 20 in some cases \citep{DEugenio+2025e,Ji+2025_local}.
A few cases where multiple lines can be fitted independently show different line widths between different broad hydrogen lines \citep{Brazzini2025,Torralba+2025b} with high significance \citetext{see \citealp{Torralba+2025b} for the role of faint \feii emission altering the
profile shape of broad H$\beta$}. These different line widths rule out a simple electron-scattering 
scenario \citep{Brazzini2025}, requiring either some role of virial broadening, turbulence, or more complex 
radiative transfer \citep[including multi-level absorption and emission;][]{Chang+2025}.
While previous studies have focused on highlighting exponential wings for the broad lines of LRDs, we have shown here that also in the blue AGN GS-3073 the broad hydrogen lines can be successfully modelled using exponential wings.
%; the double-Gaussian model appears slightly favoured when simultaneously fitting also the other Balmer lines, but the difference is subtle and, as said, subject to even very small (sub-percent) calibration, flat-fielding and continuum subtraction uncertainties (Table \ref{tab:statistics}). 
This result implies that the presence of non-Gaussian wings in Balmer profiles is not a prerogative of LRDs, but may be common to LBDs as well.

To summarize, both the Blue and Red Rosetta Stones exhibit non-Gaussian broad Balmer profiles, which can be successfully modelled with exponential, double-Gaussian, Lorentzian, or broken power-law prescriptions. 
Only the first two have been extensively analysed in this work, but this is a common discussion in literature \citep[e.g., ][]{Nagao+2006, Cano-Diaz+2012, Kollatschny2013, Scholtz+2021, Santos+2025}.
It is not clear yet if the overall AGN population exhibits this non-gaussianity to some degree, but is undetected in most cases because of data limitation (in particular, low SNR), and/or because a Gaussian profile is assumed for simplicity.
To further investigate this possibility, we fit the \ha emission-line complex in the quasar composite spectrum from \cite{VandenBerk2001} (Appendix \ref{appendix-ha-fit-in-sdss-composite}), using the three different models for broad emission discussed above for the Blue Rosetta: single broad Gaussian, double broad Gaussian, and exponential scattering kernel. 
Our results are displayed in Fig.~\ref{fig:vandenberk+2001_composite_template_fits}. 
Similarly to what retrieved for both the Red \citep{Brazzini2025} and Blue Rosettas, a single Gaussian is ruled out, while both the double Gaussian and exponential models fit the data quite well (a more thorough investigation in individual SDSS quasars will be presented in a separate paper; Trefoloni et al. in prep.).
Therefore, we can conclude that extended, non-Gaussian wings in hydrogen lines may indeed be present not only in LRDs and LBDs, but also in `standard' quasars.
The opposite is not necessarily true -- not \textit{all} LRDs exhibit exponential profiles, given the difficulty of simultaneously modelling multiple lines, the limits imposed by SNR, and some recent works apparently preferring the double Gaussian model \citep[][ Schotlz et al. in prep.]{Deugenio+2025_irony}.

Differing from GN-28074, \ha and \hb in GS-3073 exhibit similar broad-line widths, and are indeed modelled with a single set of parameters, but it is currently unknown if consistent Balmer-line widths are a feature of all LBDs. We do not compare the width of broad \hb in the quasar stack, due to the additional complication of broad \feiiperm emission. 
For LRDs the picture is also unclear, since 
there are both systems which exhibit different line widths \citep{Brazzini2025,Torralba+2025b}, and 
systems with consistent line widths \citep{Deugenio+2025_irony,Nikopoulos+2025}.

Despite high-quality, high-resolution observations \citep{Ubler+2023}, GS-3073 presents no Balmer (or any other) absorption. In contrast, the incidence of absorption in LRDs is markedly higher than in normal AGN \citep[10\%; e.g.,][]{Matthee+2024,Maiolino+2024}. 
The current fraction is almost certainly under-estimated, due to SNR and spectral-resolution constraints \citep[e.g.,][]{DEugenio+2025e}.
%while many LRDs show
These absorbers display varying degrees of kinematic complexity \citep[e.g.,][]{DEugenio+2025e,Deugenio+2025_irony,Naidu+2025}. 
In this context, the lack of Balmer absorption in Blue Rosetta
is in line with
the bulk of local quasars, and suggests an un-obstructed view towards the accretion disc.

In terms of emission-line ratios, LRDs present remarkably high Balmer decrements in their broad lines
\citep{Juodzbalis+2024,Brooks+2025,DEugenio+2025c,DEugenio+2025e,Deugenio+2025_irony,Torralba+2025b}, and large Paschen-to-Balmer ratios as well
\citep{DeGraaff+2025b}. At the same time,
the narrow lines tend to present very low dust attenuation
\citep{DEugenio+2025c,Lin+2025_local,Ji+2025_local,Deugenio+2025_irony,Nikopoulos+2025},
although the presence of Balmer absorption near rest-frame wavelengths can significantly alter the observed decrement \citep{DEugenio+2025e,Deugenio+2025_irony}, particularly in medium-resolution observations, which can completely hide even substantial absorption \citep{DEugenio+2025e}.

Together, the large decrement of the broad lines and the small decrement of the narrow lines have been interpreted as evidence for localized dust near the BLR in the AGN \citep[e.g.,][]{Nikopoulos+2025}, or the result of collisional excitation as the primary emission mechanism of the broad lines \citep[e.g.,][]{Ji+2025_local,Torralba+2025b,Torralba+2025a}.
Case-B recombination ratios combined with known dust-attenuation curves have been ruled out in one case \citep{Deugenio+2025_irony,Nikopoulos+2025}, based on the line ratios of broad \ha, \hb, and \hg. However, observations with the necessary SNR are very few, hence it is unclear how this result generalizes to the wider population.
It is clear, nevertheless, that the Blue Rosetta and quasars in general have Balmer decrements much closer to the Case-B values \citetext{see \citealp{Ubler+2023} for
GS-3073, while \citealp{Dong+2008} find a \ha/\hb ratio of 3.06 and 10\% scatter for SDSS Seyfert 1 and quasars}. However, the apparent agreement between the observed \ha/\hb ratios of blue AGN and the Case-B value may be deceptive, since these ratios may occur even in optically thick regimes \citep{Ruff+2012}, a scenario supported by several lines of evidence \citep[including the low
Pa$\alpha$/Pa$\beta$ flux ratio;][]{Glikman+2006,Ruff+2012}.
A key difference is that the \ha/Pa$\beta$ ratios in LRDs span a broad range of 3--12 \citep{DeGraaff+2025b}, while the BLR models of \citet{Ruff+2012} remain confined to \ha/Pa$\beta \gtrsim 10$.

\begin{figure}
    \centering
    \includegraphics[width=0.8\linewidth]{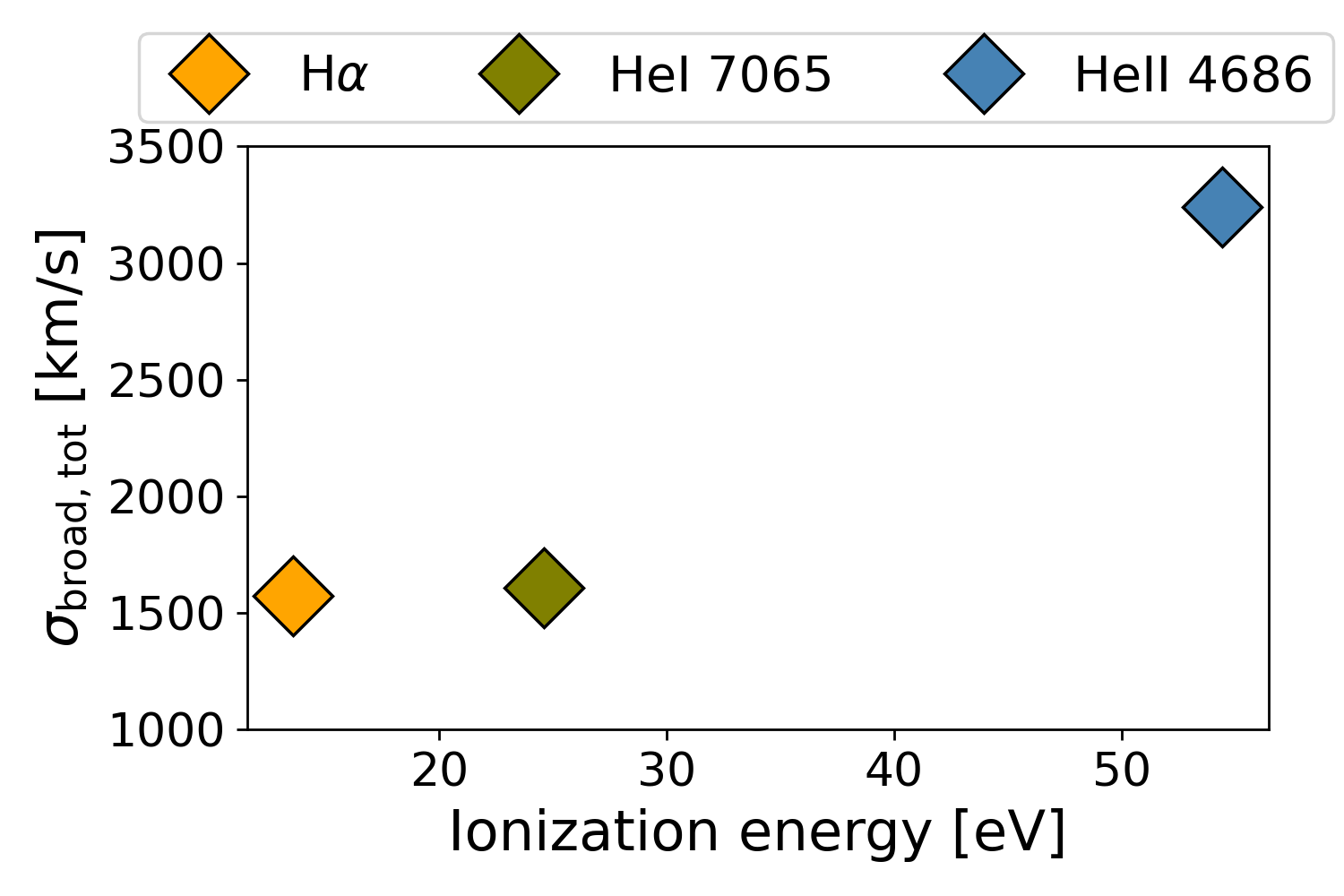}
    \caption{Correlation between line widths and ionization potentials for the \ha, \hei 7065 and \heii 4686 lines in Blue Rosetta, GS-3073. Higher ionization lines likely originate from the inner regions of the BLR, where gas clouds move at higher velocities and therefore exhibit broader profiles. }
    \label{fig:linewidths-vs-Eion}
\end{figure}

\subsection{Exponential wings and BLR stratification}

Focusing on the Blue Rosetta Stone,
we observe that the broad component of \heii is much broader than the broad components of both Balmer and \hei lines, as illustrated in Fig. \ref{fig:linewidths-vs-Eion}.
This trend supports a scenario where lines with higher ionization potentials originate from the inner regions of the BLR, which receive higher ionizing flux, and where gas clouds in virial equilibrium move faster. 
This interpretation is consistent with the BLR ionization stratification inferred from reverberation mapping studies of \heii and H$\beta$, which show that the bulk of \heii emission is much closer to the ionizing source than the Balmer-line emission \citep{Bentz2023, McDougall2025, xi2025}.
It also supports the use of virial relations for estimating the black hole mass. On the other hand, it calls into question interpretations that attribute the exponential wings of broad-line profiles to electron scattering in a putative cocoon surrounding the BLR. Such a scenario would predict similar widths for all broad emission lines; therefore, explaining the broader \heii wings would imply some sort of fine tuning, where the broadening of the Balmer lines is dominated by electron scattering, while the broadening of \heii is dominated by virial motions.

\begin{figure*}[h!]
    \centering
    \includegraphics[width=\linewidth]{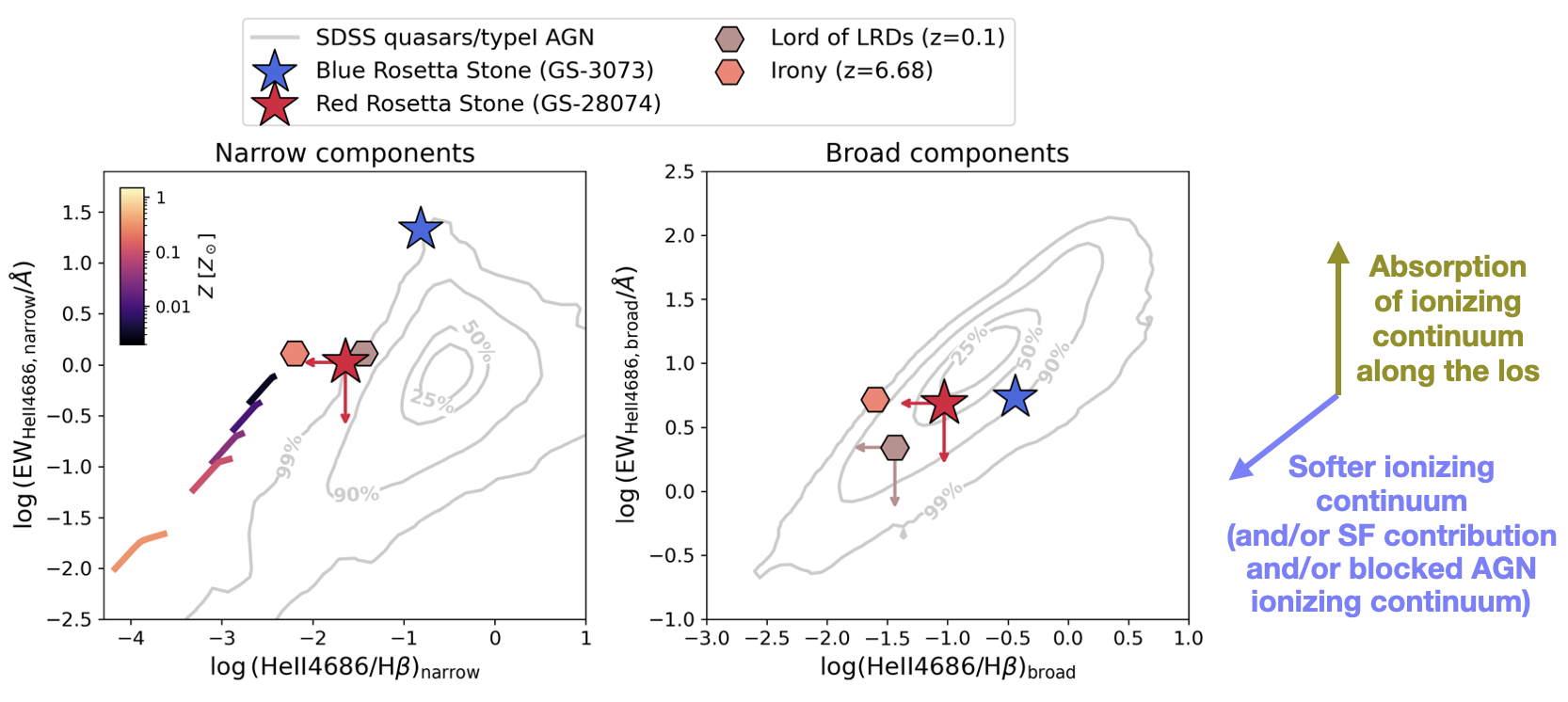}
    \caption{
    $\log (\rm EW _{HeII4686} / \text{\AA})$ versus $\log$(\heii4686/\hb) for narrow \heii emission (left panel), encompassing both the narrow and outflow components, and for broad \heii emission (right panel). 
    In both panels, the LBD GS-3073 is reported as a blue star, flanked with some significant LRDs from the literature: GN-28074 (upper limits), in red, the local Lord of LRDs at $z=0.1$, and Irony at $z=6.68$. The gray contours refer to local broad-line AGN from the SDSS DR16 quasar catalogue \citep{wu2022}. 
    In the left panel for the narrow components, we also show some predictions from photoionization models of single-burst, 1-Myr old stellar populations, with metallicities ranging from 0.001$Z_\odot$ to solar values.
    Assuming GS-3073 as representative of the LBDs population, we observe that both LRDs and LBDs occupy a different position in the $\log (\rm EW _{HeII4686, narrow} / \text{\AA})$ versus $\log$(\heii4686/\hb)$_{\rm narrow}$ plane with respect to standard AGN. 
    Instead, the difference between LRDs, LBDs and standard type I AGN  in their broad \heii emission is less evident, with LRDs and LBDs characterized by lower and higher (\heii4686/\hb)$_{\rm broad}$ values, respectively, and similar EW$_{\rm HeII 4686, broad}$ and (\heii4686/\hb)$_{\rm broad}$, both still lying in the same region of local broad-line AGN. }
    \label{fig:logEW_vs_flux}
\end{figure*}

\subsection{\texorpdfstring{\oiperm[]}{O I} fluorescent emission and \texorpdfstring{Lyman\,$\beta$\ }{Lyman Beta}% Balmer
pumping}

The scenario of an ultra-dense inner region where radiative transfer effects are important is independently supported by the detection of broad \oiperm.
For LRDs, the detection of \oiperm has been interpreted as evidence for Ly$\beta$ fluorescence \citep{Juodzbalis+2024,Tripodi+2025,Kokorev+2025,DeGraaff+2025b}.
This hypothesis is particularly compelling given that the only known LRD with both \oiperm[$\lambda 1302$] and \oiperm has a photon-flux ratio close to unity \citep{Tripodi+2025}, while other high-excitation \oiperm[] transitions remain either weak or undetected -- a natural consequence of fluorescence.
Nevertheless, \oiperm is also regularly detected in Seyfert 1 galaxies and
quasars, and indeed it is clearly seen in the stack of normal quasars shown in Fig.~\ref{fig:prism-comparison-spectra}. Therefore, the presence of \oiperm[] fluorescence is not a peculiarity of LRDs, but a general property of AGN, simply tracing ultra-dense gas in, or near the BLR.
Still, some problems with the fluorescence scenario remain -- even in LRDs. This is underscored by 
the \oiperm[1.1287~$\mu$m] -- \oiperm photon ratios being inconsistent with unity in the highest-quality observations to date \citep[][their fig.~2]{Lin+2025_local} and also by
the different line widths between \oiperm and \ha \citetext{e.g., \citealp{Juodzbalis+2024} 
, and \citealp{Lin+2025_local}, their fig.~2}. The latter issue indicates that \oiperm[] fluorescence is arising in regions farther from the black hole relative to the bulk of the gas emitting \ha; it also does not support the scenario in which \ha is primarily produced by Lyman pumping.%, as suggested by some studies.

Unfortunately, the spectrum of the Blue Rosetta Stone does not cover \oiperm, so we cannot investigate this feature. However, this will be discussed in a separate paper presenting a more extensive compilation of LBDs (Geris et al. in prep.).

\subsection{Low-ionization iron emission}

Classical type I AGN and quasars generally display clear broad \feiiperm emission \citep[e.g.,][]{Baldwin+2004,Kovacevic+2010}, seen also in the 
 \citet{VandenBerk2001} stack \citep[see e.g.][for \feiiperm deficient quasars]{boroson+green1992}. In contrast, JWST-discovered AGN display
no evidence for this FeII `pseudo-continuum' \citep{Trefoloni+2025}, completely missing
the strong 4464\AA--4684\AA\ bump, even in extremely deep observations of Blue Rosetta Stone. Intriguingly, an indication of the 2200\AA--3500\AA\ \feiiperm bump is tentatively seen in GN-z11 (another LBD at z=10.6; \citealp{Ji+2025_z11,Maiolino+24b}; but see also \citealp{Alvarez-Marques2025,Crespo2025} for a different view in this source), but with a spectral shape different than in classical AGN, possibly indicating different physical conditions and different turbulence properties.
Surprisingly, \feiiperm is regularly detected in LRDs, although the emission arises from forbidden transitions and is much narrower than what seen in quasars \citep{Lin+2025_local,Ji+2025_local,Lambrides+2025,Deugenio+2025_irony,Kokorev+2025}. A proposed explanation is that it arises from a cool envelope \citep{Lin+2025_local,Torralba+2025b}, although other solutions cannot be ruled out \citep[see e.g.,][]{Ji+2025_local}.

\begin{figure}
    \centering
    \includegraphics[width=0.95\linewidth]{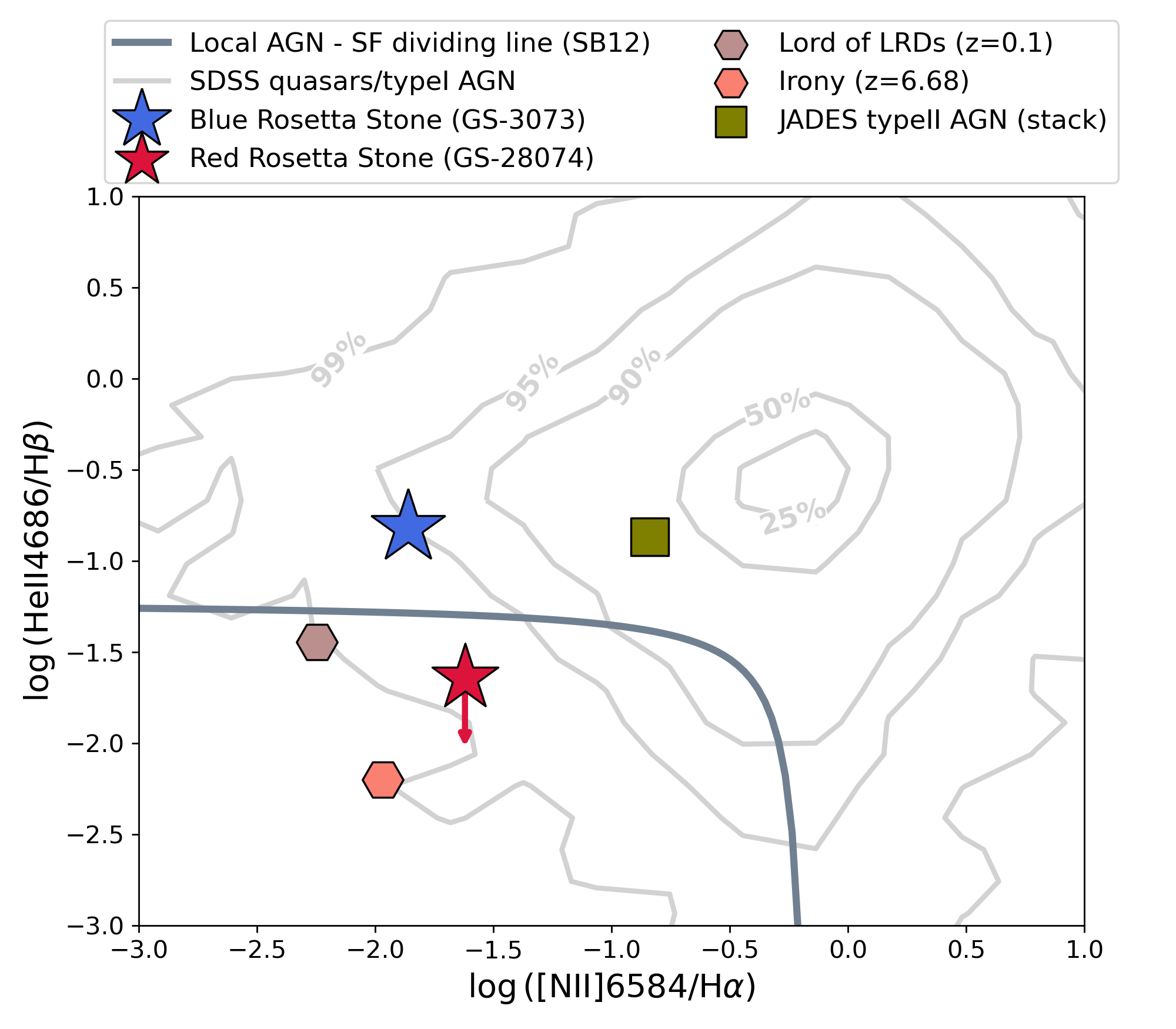}
    \caption{$\log$(\heii4686/\hb) versus $\log$(\nii6584/\ha) for GS-3073 (blue star), flanked with some typical LRDs: GN-28074 (red star), Lord of LRDs, and Irony. The gray contours are local broad-line AGN from the SDSS DR16 quasar catalogue \citep{wu2022}. The green square represents the stack of type II AGN from JADES \citep{Scholtz+2025_typeIIsample}.
    GS-3073 lies in the AGN region, with $\log$(\heii4686/\hb)~$=-0.82 \pm 0.09$ and $\log$(\nii6584/\ha)~$= -1.86 \pm 0.10$, slightly shifted towards lower $\log$(\nii6584/\ha) values with respect to the bulk of SDSS AGN -- probably a metallicity effect. 
    Instead, LRDs at all redshifts lie in the star-formation region, covering significantly lower values of both $\log$(\heii4686/\hb) and $\log$(\nii6584/\ha), probably due to a combined effect of low metallicity and absorption of ionizing photons from the dense gas envelope. }
    \label{fig:logHe2_vs_N2}
\end{figure}

\subsection{High-ionization lines: \texorpdfstring{\heii}{He II} $\lambda 4686$}

The weakness of high-ionization lines in JWST-selected AGN and LRDs has already been pointed out in previous studies \citep{Juodzbalis+2025, Zucchi+2025, Lambrides+2024, Lambrides+2025}. 
In agreement with these findings, the Red Rosetta Stone does not show evidence for \heii, as already mentioned in \cite{Juodzbalis+2025}, while the Blue Rosetta Stone shows clear detection of \heii$\lambda4686$, both narrow and broad (left panels of Fig. \ref{fig:gs3073-linefits}) -- we discuss in the following whether this detection is due to the high brightness and high SNR of its spectrum, or it is stronger than in other populations of AGN.

From our emission-line fits,
% integration,
 we measure a \heii rest-frame equivalent width (EW) of $21 \pm 3$ \AA\ for the narrow emission, encompassing both the narrow and outflow Gaussian components, consistent with the previous estimate reported by \cite{Ubler+2023}. 
The broad emission instead exhibits an EW of $36 \pm 7$ \AA.
The large errorbars on the EW are due to the large measurement uncertainties % errors
in the line and continuum estimates.

In Fig.~\ref{fig:logEW_vs_flux}, we compare the location of Blue Rosetta Stone on the log(EW${_\textrm{\heii}}$) versus log(\heii/\hb) diagram, which
relate the line-to-continuum strength (e.g., covering factor, optical depth, geometry) to the hardness of the ionizing spectrum. 
We separate the lines in narrow+outflow and broad components, and use local quasars from the SDSS DR16 quasar catalogue as a benchmark \citep{wu2022} (grey contours). 
We also compare with 
the Red Rosetta, for which we show the 3$\sigma$ upper limits (assuming line widths of $150$~ km/s and $6000$~ km/s, respectively), and with two additional
 LRDs from the literature: the `Lord of LRDs' J1025+1402, a recently-discovered local LRD at $z = 0.1$ (\citealp{Ji+2025_local}, see also \citealp{Lin+2025_local}) and RUBIES-EGS-49140, dubbed `Irony', currently the only high-redshift ($z=6.68$) LRD exhibiting \heii$\lambda 4686$ detection \citep{Wang+2025_irony, Deugenio+2025_irony}.
 For the narrow components, we further compare with the predictions from \textsc{Cloudy} \citep[v23.01;][]{Ferland+1998,Gunasekera+2023} photoionization models of star-forming galaxies. We used \textsc{BPASS} v2.3 simple stellar population models \citep{byrne+2022}, assuming a single starburst episode 1~Myr ago and a
 \citet{Chabrier2003} initial mass function. The clouds assume a single density of $\log\,n_\mathrm{H} = 10^3~\rm cm^{-3}$ and solar-scaled abundances.
 Different metallicity regimes are explored, ranging from $0.001 Z_\odot$ to solar values. Each line represents a model with fixed metallicity, but with varying ionization parameter (increasing left to right from $\log\,U=-4$ to $-1$).
Our EW$_\textrm{\heii}$ also includes the transmitted emission line from the input stellar spectrum, since this cannot be reliably distinguished from the nebular component at the resolution of the prism \citep[also, for the lowest metallicities relevant here, stellar \heii$ \lambda 4686$ may be significantly narrower than in more metal-rich young stars; e.g.,][]{Berg+2025}.
% and GN-28074, where \heii is not detected but for which we report the $3 \sigma$ upper limits in \heii flux and EW in both narrow and broad components, assuming line widths of $150$~ km/s and $6000$~ km/s, respectively. 
% While the former assumes that the kinematics of narrow \heii is tied to that of narrow \hb, for the broad component the choice of line width is not straightforward. In fact, as we have shown (Fig. \ref{fig:linewidths-vs-Eion}), the width of broad \heii may significantly differ from broad Balmer emission. 
% Therefore, we measure the \heii EW by assuming: 1) $\textrm{FWHM}_\textrm{\heii, broad}=\textrm{FWHM}_\textrm{\hb, broad}$; 2) $\textrm{FWHM}_\textrm{\heii, broad}=1.9 \cdot\textrm{FWHM}_\textrm{\hb, broad}$, where the scaling factor is the ratio of the broad \heii over \hb FWHMs for GS-3073. 
% We then adopt the most conservative result, that is, the broader width estimate. 

We note that for the broad component, both Blue Rosetta and Irony (an LRD) lie within the distribution of normal quasars. Even though they are
near the outer (90\textsuperscript{th}-percentile) density contour, their location is overall consistent with an AGN-like ionizing spectrum, although possibly under slightly different physical conditions (density or ionization parameter).
Interestingly, despite their brightness, the upper limits on both Red Rosetta and Lord of LRDs are not very constraining in terms of \heii emission, and could very well be compatible with normal AGN. Yet, we note that these are 3$\sigma$ upper limits; tighter upper limits pushing towards the bottom-left corner of the diagram would favour soft ionizing continua.

As for the narrow emission lines on the same diagram (panel on the left in Fig.\ref{fig:logEW_vs_flux}), the situation looks more peculiar, whereby both Blue Rosetta and the LRDs are offset relative to the local quasar distribution towards the upper-left region of the diagram, i.e. either higher EW$_\textrm{\heii}$ and/or lower \heii/\hb, although in the case of Red Rosetta we only have upper limits. An offset towards lower \heii/\hb could be due to either an intrinsically softer ionizing continuum, and/or contribution by star formation in the host. It could also result from a gaseous medium blocking the AGN ionizing radiation, or simply by BLR clouds with large covering factor. Higher EW($_\textrm{\heii}$) would require either intrinsically stronger line emission, or a suppression of the AGN ionizing continuum along our line of sight, resulting in a continuum that is dominated by host-galaxy emission or reflected AGN light.

\begin{figure*}
    \centering
    \includegraphics[width=0.8\linewidth]{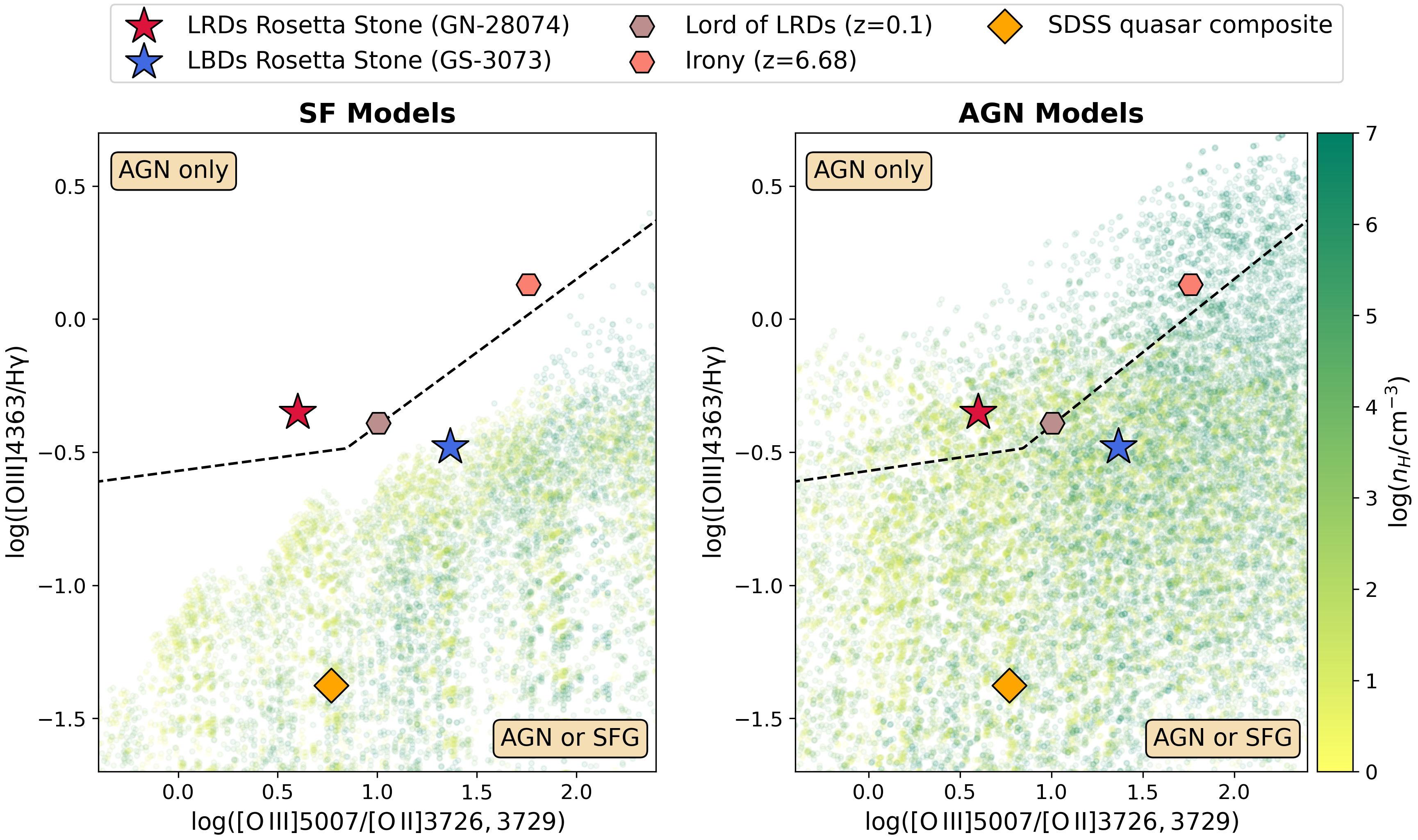}
    \caption{Auroral-\oiii diagnostic diagram from \cite{Mazzolari+2024_newdiag} for GS-3073 (blue star), flanked with LRDs: GN-28074 (red star), Lord of LRDs, and Irony. Errors on the Blue Rosetta's flux ratios are of few \%, too small to be reported in the figure. 
    The SDSS  composite quasar spectrum from \cite{VandenBerk2001} is reported for comparison. Values from the large grid of \textsc{Cloudy} models \citep[drawn from the parent HOMERUN grid;][]{Marconi+2024} are coloured by hydrogen density, with star-forming models in the left column
    and AGN models in the right column.}
    \label{fig:auroral_diagnostic}
\end{figure*}

We then locate the same sources on the  $\log$(\heii$\lambda4686$/\hb) versus $\log$(\nii$\lambda6583$/\ha) diagram (Fig.~\ref{fig:logHe2_vs_N2}), again flanked by the SDSS DR16 quasar catalogue. We also plot the location of the JADES type 2 AGN stack \citep[from][]{Scholtz+2025_typeIIsample} with a squared symbol.  The solid line indicates the dividing line between AGN and SF identified by \cite{Shirazi+2012}, and tested more extensively by \cite{Tozzi+2023} on local galaxies.
Blue Rosetta lies in the AGN region, although shifted towards lower N2 values with respect to the bulk of local quasars, likely because of low-metallicity effects in the narrow-line region. 
Red Rosetta and the other two LRDs instead lie in the star-formation region.
However, if it were not for low [NII]/H$\alpha$ ratio, which might be simply a consequence of low metallicity, they would probably lie within the distribution of SDSS AGN.

Summarizing, in terms of broad \heii, Blue Rosetta is clearly more prominent than Red Rosetta and the other LRDs, indicating that the BLR is seeing harder and softer ionizing continuum in the two cases, respectively. 
This can be an effect of different incident SEDs (with the Blue Rosetta exhibiting an intrinsically 'hotter' SED), but also of different BLR properties: for instance,
%Yet, we cannot exclude also an effect of ionization parameter: 
higher density of the BLR clouds in LRDs could result into lower ionization parameter, hence suppression of \heii emission - this effect will be explored in a separate paper.
% \RMcomm{Unless Xihan has time to check this scenario with Cloudy}. 
Yet, we emphasize that both Blue Rosetta and the LRDs (even their upper limits) are well within the distribution of quasars, hence there is yet not strong evidence of prominent deviations from normal quasars. The offset in terms of the narrow lines is more challenging to explain: blocking the nuclear radiation should move points towards the locus of SF galaxies. The scenario whereby the nuclear AGN source absorbed along our line (or strongly anisotropic) of sight might work for Blue Rosetta - this is essentially the puffed up accretion disc scenario in the super-Eddington model \citep[e.g.][]{Madau+2024,Madau2025,King2025,Zucchi+2025}.
Yet, the latter scenario would not work well for Irony and Lord of LRDs, whose EW$_\textrm{\heii}$ of the broad component is fairly consistent with normal quasars. Therefore this explanation is not completely satisfactory for the whole class of AGN revealed by JWST.

\subsection{Auroral lines}

Both Blue and Red Rosetta Stones show prominent \oiii$\lambda4363$ auroral line emission. This allows us to locate them in the new diagnostic diagram proposed by \cite{Mazzolari+2024_newdiag}, featuring \oiii$\lambda4363$/H$\gamma$ vs \oiii$\lambda5007$/\oii$\lambda3727$. This diagram is shown in Fig.\ref{fig:auroral_diagnostic}.
\cite{Mazzolari+2024_newdiag} noticed that the region of the diagram with high \oiii$\lambda4363$/H$\gamma$ and low \oiii$\lambda5007$/\oii$\lambda3727$ is populated only by AGN, both in the local Universe and at high redshift -- while the rest of the diagram at lower \oiii$\lambda4363$/H$\gamma$ ratios can be populated by both AGN and SF galaxies. 
They interpreted this as due primarily to the AGN X-ray heating being more effective in heating the ISM at a given ionization parameter (traced by \oiii/\oii), hence boosting the auroral line emission. 
This is confirmed by photoionization models, both in \cite{Mazzolari+2024_newdiag} and in the large grid of \textsc{Cloudy} models presented in \cite{Jones+2025_dormant} and drawn from the parent HOMERUN grid by \cite{Marconi+2024}. 
We show the models presented in the latter paper in Fig.\ref{fig:auroral_diagnostic}, where the left panel shows the star forming (SF) models, while the right panel shows the AGN-only models. Models span a broad range of ionization parameters and densities, and are colour coded by gas density. 

We show on the same diagrams the location of Blue and Red Rosetta Stones. 
Clearly, SF models fail to reproduce very high values of \oiii$\lambda4363$/\hg for Red Rosetta (for which log(\oiii$\lambda4363$/\hg)$_{\textrm{GN-28074}}= -0.350 \pm 0.014$, and log(\oiii$\lambda5007$/\oii$\lambda 3727)_{\textrm{GN-28074}}= 0.60 \pm 0.08$), despite models reaching extremely high densities (up to $10^7~\text{cm}^{-3}$, much higher than the critical density of \oiii$ \lambda 5007$), while AGN can reproduce fairly high values of this ratio.
Paradoxically, Blue Rosetta, which displays prominent \heii, is located in the SF-AGN mixed region of the diagram (log(\oiii$\lambda4363$/\hg)$_{\textrm{GS-3073}}=  -0.49 \pm 0.02 $, and log(\oiii$\lambda5007$/\oii$\lambda 3727)_{\textrm{GS-3073}}= 1.362 \pm 0.013 $), while Red Rosetta is solidly in the AGN-only region of the diagram. Note that invoking densities even higher than those explored in the models presented by \citet{Jones+2025_dormant} would not help, as they are approaching the critical density of \oiii$\lambda 4363$ ($5\times 10^7~\text{cm}^{-3}$), where this line starts to saturate relatively to \hg.
Moreover, significantly higher densities would place the emitting gas in the BLR realm, in which case the \oiii$\lambda 4363$ line should be considerably broader. Therefore, a more likely scenario is that the medium emitting \oiii$\lambda 4363$ is irradiated by X-rays, which are effective in heating it and making it fairly hot. 
The presence of AGN may appear in contrast with the absence of \heii in Red Rosetta. However, it is also true that AGN accreting well below Eddington are characterized by prominent X-ray emission and modest blue bump. This would be in tension with the scenario that LRDs are super-Eddington sources \citep{Madau+2024, Lambrides+2024, Pacucci+2024, Inayoshi2025};
%\RMcomm{to me mostly due to yet another massive bandwagon effect}\fde{One could also add that sub-Eddington accretion in low-mass, young galaxies is naturally conducive of amassing large amounts of gas: the cool accretion rate is high, and feedback is anaemic}\fde{If we are bold, we could make a parallel with stochastic star formation: just like galaxies undergo bursts of star formation followed by quenching, so AGN colud undergo cycles of high and low accretion states; if the AGN was unable to clear the BH surrounding completely, then re-accretion could start a low accretion LRD phase}; 
Within this context and in the same direction, we note that the only LRD for which the BH mass has been measured directly (dynamically), A2744-QSO1, is actually found to be accreting well below Eddington -- $L_{\rm AGN}\sim 10^{-2}L_{\rm Edd}$ \citep{Juodzbalis+2025b}.
Of course, given that these properties are being assessed only for a small subset of very well studied objects, it is possible that LRDs are a heterogenous population with different properties.
% Therefore, these results suggest that one may be more open to these scenarios that are orthogonal to the commonly endorsed opinion about LRDs.

\begin{figure}
\centering
\includegraphics[width=\linewidth]{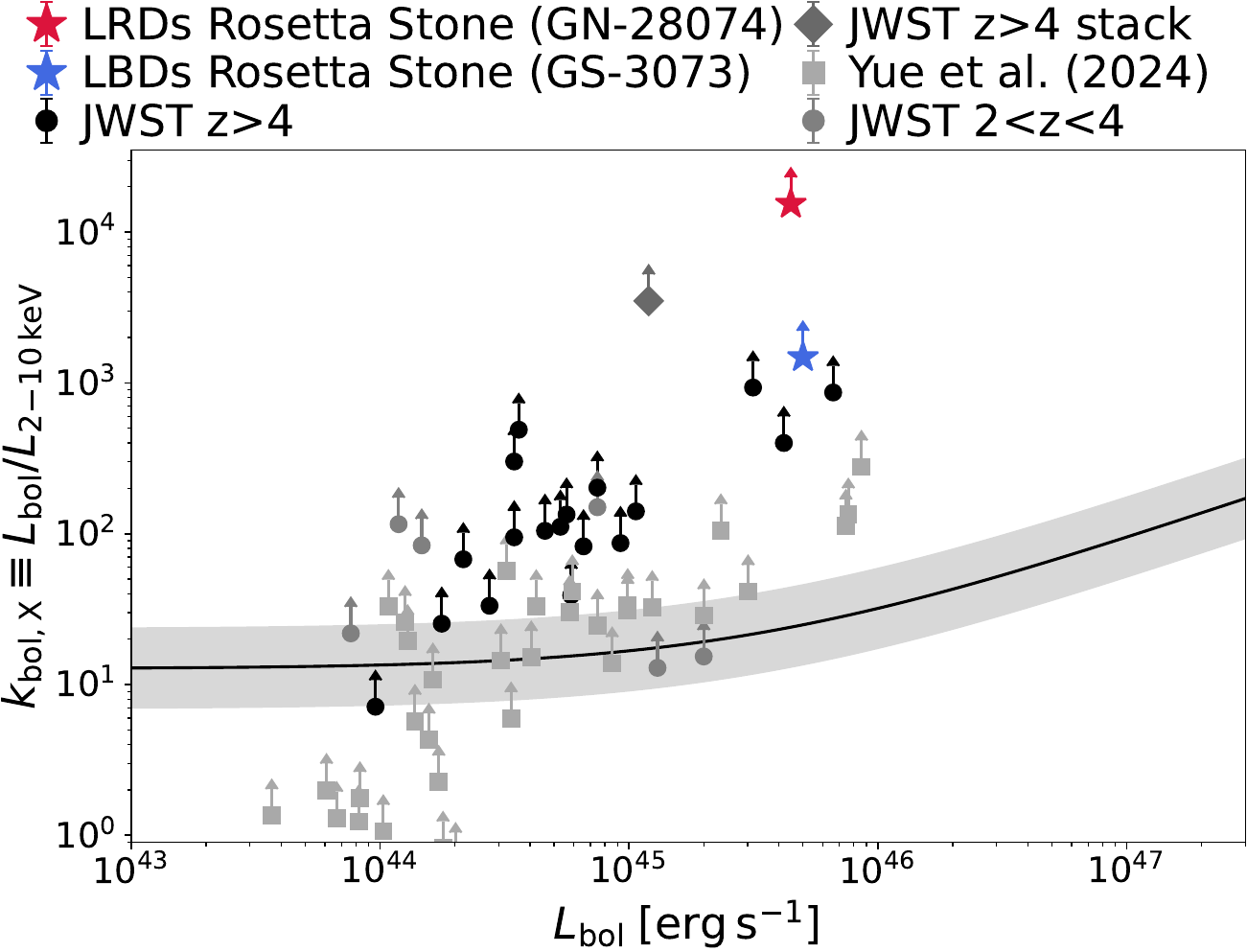}
\caption{Ratio $k_\mathrm{bol,X}$ between AGN bolometric luminosity and the X-ray (2–10 KeV) luminosity, as a function of bolometric luminosity. Both Blue and Red Rosetta stones stand out as extremely X-ray weak sources, lying over an order of magnitude above the local $L_\mathrm{bol}\text{--}k_\mathrm{bol,X}$ relation
\citep[black line, with the grey shaded region representing 1-$\sigma$ scatter;][]{Duras2020}. While remaining undetected, other individual JWST AGN do not have constraining lower limits \citetext{grey circles, \citealp{Maiolino+2025}; grey squares \citealp{Yue+2024}}. However, when stacked, their combined lower limits become extremely X-ray weak \citep[grey diamond;][]{Maiolino+2025}.
Note that even using an order of magnitude lower bolometric correction, Red Rosetta would still remain X-ray weak. X-ray weakness is a common feature of LBDs and LRDs.
}
\label{fig:xray}
\end{figure}

\begin{figure}
\centering
\includegraphics[width=\linewidth]{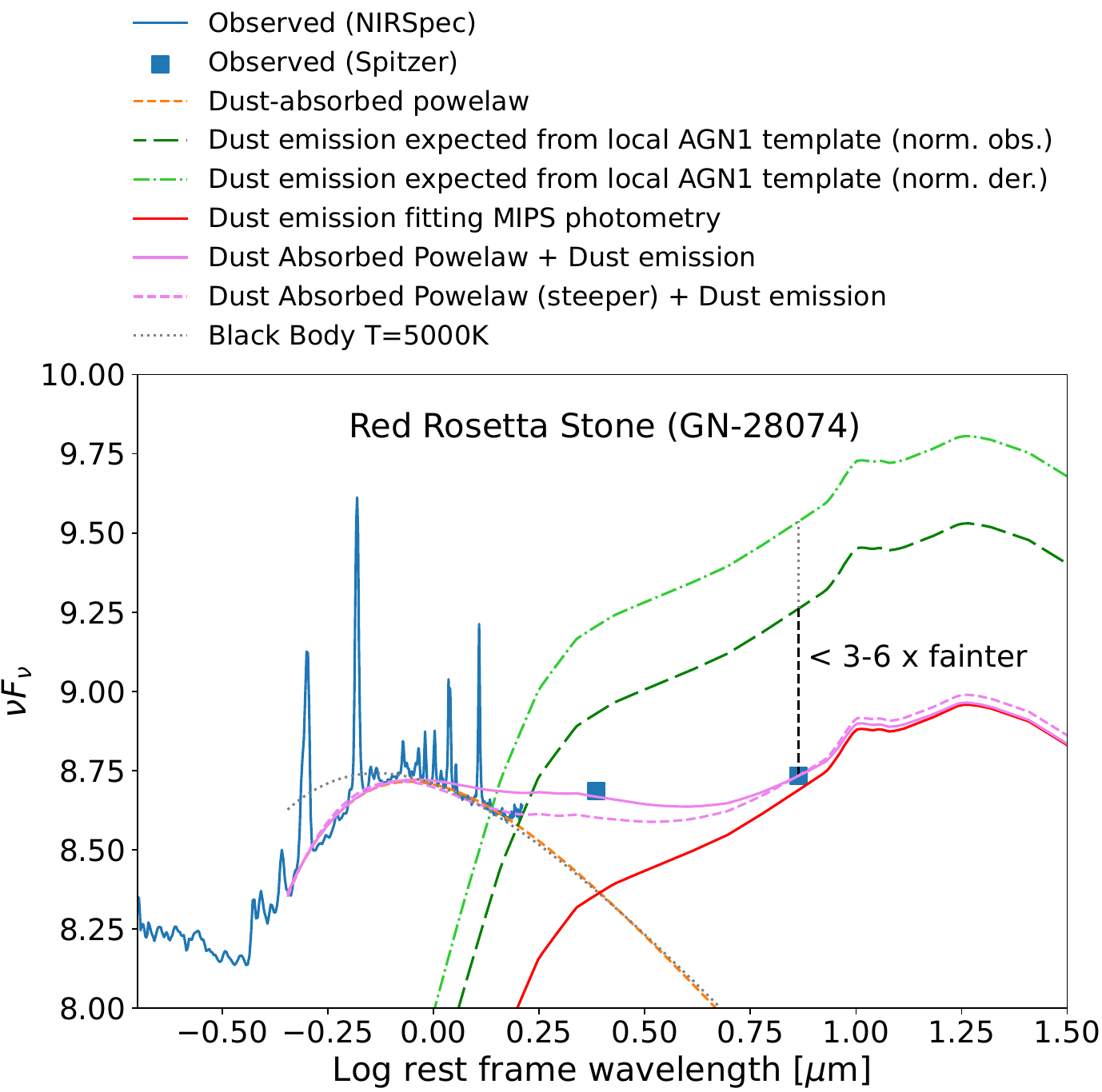}
\caption{Fit and analysis of the optical and mid-IR spectral energy distribution of Red Rosetta and, specifically, its NIRSpec prism spectrum (blue solid) and the IRAC-Ch4 and MIPS photometric points (blue squares). The orange short dashed line is a dust-absorbed powerlaw fit to the optical spectrum. The green long-dashed and dot-dashed spectra show the expected hot dust emission based on local type I AGN templates, with different normalization (see text). The red solid line shows the local torus template scaled to match the observed MIPS point, and the violet solid line show its combination with the reddened power law. The short-dashed violet line shows the same combination where a steeper powerlaw is adopted for the optical spectrum. The gray dotted line is a fit to the red part of the spectrum with a 5000 K blackbody.}
\label{fig:red-rosetta-sedfit}
\end{figure}

\begin{figure}
\centering
\includegraphics[width=\linewidth]{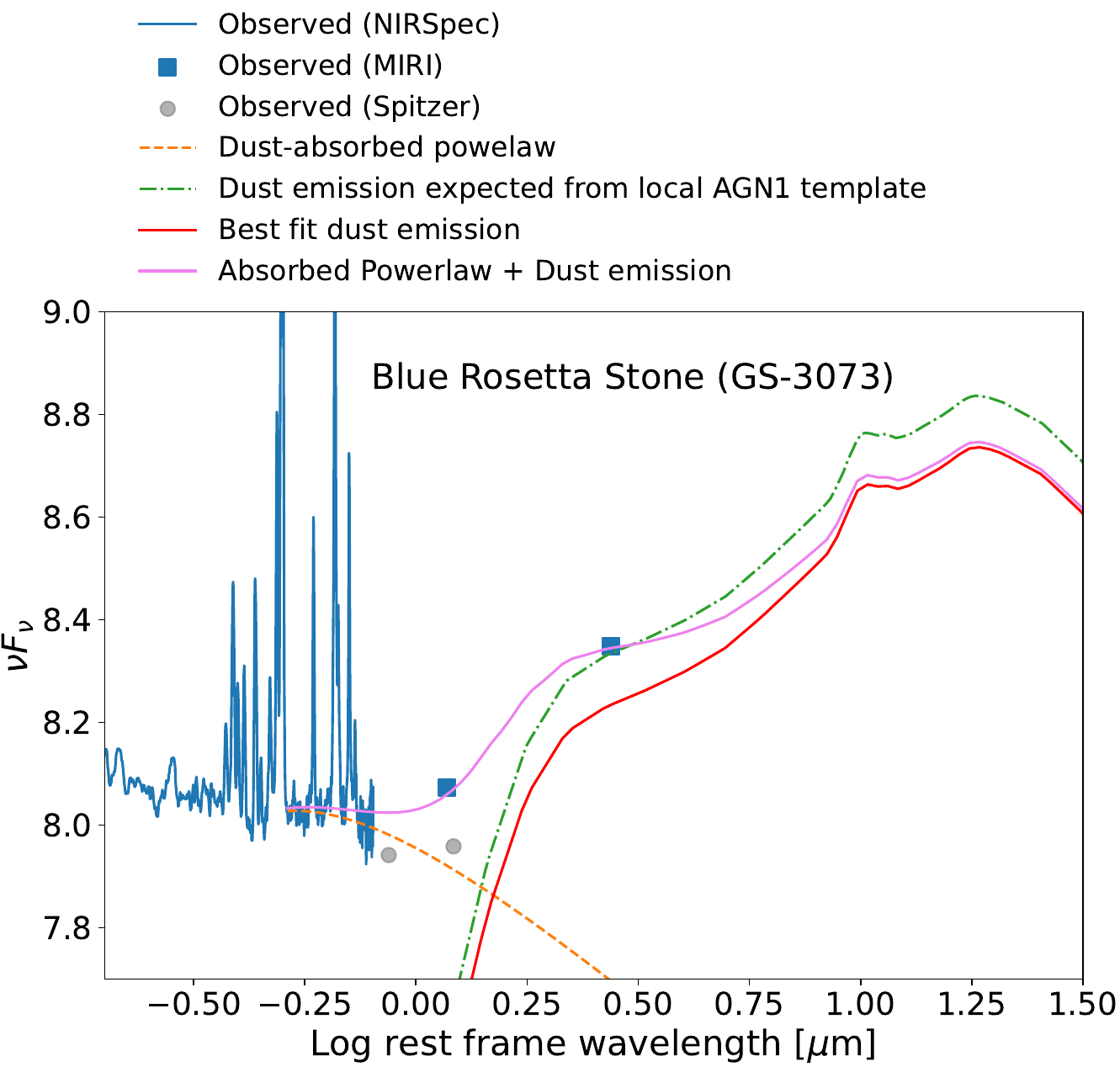}
\caption{Fit and analysis of the optical and mid-IR spectral energy distribution of Blue Rosetta and, specifically, its NIRSpec prism spectrum (blue solid) and the MIRI photometric points (blue squares); Spitzer points are also shown (orange circles). The orange short dashed line is a dust-absorbed powerlaw fit to the optical spectrum. The green dot-dashed line indicates the hot dust emission expected from local type I AGN templates. The red solid is the same template rescaled to fit the MIRI photometric points, and the violet solid line is its combination with the reddened power law.}
\label{fig:blue-rosetta-sedfit}
\end{figure}

\subsection{X-ray emission}
\label{section-X-ray}

As already discussed in the previous sections,
both Red and Blue Rosetta Stones are observationally extremely X-ray weak, with neither source detected even in the deepest Chandra observations (Fig.~\ref{fig:xray}).
The stringent upper limits on their X-ray emission place them among the AGN with the highest $k_{\text{bol}}$ ever reported.

In the case of the Red Rosetta, the prominent Balmer absorption lines indicate that we are seeing the nucleus through dense absorbing gas, and this may well be responsible for extreme (Compton thick) absorption of the X-ray emission coming from the nuclear region, as suggested by varioius groups \citep[e.g.][]{Maiolino+2025,Rusakov+2025}. In this scenario, the LRDs would be intrinsically X-ray luminous, and the X-ray emission could heat the gas that primarily produces the auroral line emission, although the lines of sight reaching such gas should not be Compton thick.

In contrast, the absence of X-ray emission in the Blue Rosetta Stone is more puzzling. 
Indeed, in this case there is no evidence for dense gas along the line of sight, as no Balmer absorption is detected. 
While the presence of absorbing material cannot be entirely ruled out, as observed in some local type I AGN \citep[e.g.][]{Risaliti2009,Risaliti2011,Maiolino2010,Kumar2024}, such gas is unlikely to be Compton thick. It is therefore possible that in LBDs the observed X-ray weakness is intrinsic or driven by an unusually soft ionizing spectrum, analogous to what is observed in local narrow-line Seyfert I galaxies, possibly due to a weak or cold corona, or strongly anisotropic radiation \citep{Maiolino+2025,Madau+2024,Madau2025,King+2025,Pacucci+2024}. If LRDs and LBDs share the same central engine, then intrinsic X-ray weakness may also apply to LRDs, although this may imply re-visiting the interpretation of the auroral line strength as due to X-ray heating in these systems. An additional possible scenario, which could apply to both LBDs and LRDs, is the one proposed by \cite{Pacucci2026} in which the inner X-ray emission would be absorbed by an envelope of dense gas, which, at the same time, emits at a temperature of 40,000~K, hence producing the blue spectrum in LBDs and the redder spectrum in LRDs via some dust reddening. 
The only potential issue in this case is that in their scenario the X-ray absorbing medium is mostly external and responsible for the Balmer absorptions seen in LRDs -- this would imply a different geometry for LBDs, or some additional tuning of the model.

\subsection{Hot dust emission}
\label{section-hot-dust}

In section \ref{section-mid-IR}, we have shown that both Red and Blue Rosetta Stones show clear evidence for hot dust emission, which is typically seen in AGN. In order to have prominent emission from dust at temperatures close to sublimation, it requires irradiation from a strong UV radiation field. This, per se, is in conflict with the Black-Hole-star, or cocoon, scenario, according to which no UV radiation should reach the circumnuclear dust. While this is a first qualitative assessment, in this section, we assess more quantitatively the prominence of the hot dust emission.

\subsubsection{Red Rosetta Stone}

In Fig.~\ref{fig:red-rosetta-sedfit} we show the NIRSpec prism spectrum of Blue Rosetta (solid blue line) and the Spitzer MIPS and IRAC-Ch4 photometric points (we do not use the other IRAC channels because heavily contaminated by emission lines). We start by fitting the continuum observed in the NIRSpec spectrum
with a dust absorbed power law. In this case we are using the extinction curve recently identified by \cite{Sun2026} at high-z via the analysis of galaxies in the background of other galaxies (these are proper extinction curves, while other studies have inferred attenuation curves); using an SMC extinction curve would result in a slightly worse fit. The fit requires an intrinsic powelaw with slope $\beta = -2.49\pm0.15$ (where $\beta$ is defined as $F_\lambda \propto \lambda ^\beta$), which is very close to the slope expected for classical accretion discs \citep[$\beta=-2.33$,][]{Lynden-Bell1969Natur,SS73}, and $A_V=2.1\pm 0.2$; interestingly, this extinction is very close to the absorption that one would infer from the Balmer broad lines ratios by simply assuming case B and the same extinction curve: $A_V=2.4-2.7$ (depending on the adopted pair of lines). Then we use the local SED template obtained by \cite{Garcia-Bernete2019} by fitting the nuclear emission of a sample of nearby type I AGN
and decomposing the dusty torus emission from the disc emission. The long dashed dark green line shows the torus emission expected from the local type I AGN template, renormalised to the 
observed 1.15$\mu$m emission - this is about a factor of three higher than the mid-IR emission observed with Spitzer-MIPS (blue square). In principle, assuming that the optical continuum is entirely coming from the reddened accretion disc,  one should renormalize to the de-reddened optical emission; however, one should do the same also for the template, where the optical continuum is the stack of different spectra with different slopes and different reddening, which makes the procedure very uncertain. Yet, with all these caveats in mind, and fitting the optical emission of the template with a reddened powerlaw, one would infer an expected hot dust emission given by the dot-dashed lighter green line, which is about six times higher than the observed MIPS photometry. Summarizing, the hot dust emission expected from the local type I AGN template is between three and six times higher than observed. However, one should consider that these excess estimates are upper limits, as they are inferred under the assumption that the optical continuum is entirely AGN-related in nature - if there is any contribution from a host galaxy \citep[as inferred by ][]{Juodzbalis+2024,Juodzbalis+2025}, then the expected mid-IR flux from the template would be reduced accordingly. One could potentially normalize the template also to the broad H$\alpha$, which is often taken as a proxy of the AGN bolometric luminosity; however, its large EW suggests that the covering factor of the BLR clouds in these objects might be much higher than in local AGN \citep{Maiolino+2025} or that it is boosted by collisional excitation or radiative transfer effects \citep[e.g.][]{Chang+2025,Deugenio+2025_irony,Torralba+2025b,Nikopoulos+2025}, which would add further complications and jeopardize its reliability.

Although the magnitude of the hot dust deficiency emission is subject to uncertainty (a factor between 3 and 6, or less), it is clear that Red Rosetta has less hot dust emission than local type I AGN (although still very prominent relative to normal star forming galaxies). Yet, this can potentially and simply be explained by the filtering of the UV emission by the same neutral gas that is responsible for the Balmer absorption lines; if the dusty medium is shielded by the same absorbing medium as along our line of sight, then this would be enough to completely remove the radiation shortward of the Lyman limit. For a standard AGN SED, this is about 83\% of the UV radiation, at wavelengths where dust is most effective in absorbing light. A suppression of the impinging radiation by a factor of $\sim 6$ would certainly be enough to suppress the resulting mid-IR emission by a similar factor. The absorption shortwards of the Balmer break would further reduce the impinging radiation. Additionally, one should consider that these high-z AGN have much lower metallicity than local templates, and therefore the dust-to-gas ratio is unavoidably much lower; this can play an additional, important role in the intensity of the hot dust emission. An example of a model in which very little dust mass (hence negligible thermal emission) can produce significant dust reddening is presented in \cite{Pacucci2026}.

The red solid line shows the local type I AGN template rescaled such that, when combined with the extrapolation of the model fitting the optical continuum, it reproduces the observed MIPS emission. The violet solid line shows the sum of the two, which also nicely reproduces the IRAC-Ch4 photometry, but it overshoots the red end of the NIRSpec spectrum. It is alternatively possible to fit the NIRSpec spectrum with an absorbed steeper powerlaw and the local hot dust template, as shown with the dashed magenta line, but in this case the IRAC-Ch4 point is under-fitted. This may indicate that the local hot dust emission template is inadequate for Red Rosetta, and that an additional hot dust (T$\sim 1000-2000~K$) component is needed. However, one should consider that deviations are of the order of 10\%-20\% and these may originate from inter-calibration problems between Spitzer and JWST, and variability may also have played a role; indications of these possibilities will be discussed further for the Blue Rosetta.

Finally, with a dotted gray line we also show the case of fitting the red part of the optical NIRSpec spectrum with a T=5000~K blackbody, as suggested by \cite{Liu2025} for their model of a spherical envelope around a black hole accreting at a rate 100 times above the Eddington limit. This model has issues in explaining the origin of the emission lines (as the ionizing radiation would be totally absorbed or, alternatively, the emission lines could not escape if produced inside the envelope); leaving aside this important caveat, the model could explain the red part of the optical continuum, while the blue part  would require some sort of stratification of the atmosphere or wavelength dependent emissivity to explain the steeply declining flux of the continuum shortward of 1$\mu$m. A more thorough discussion of these aspects is deferred to a separate paper. Here we simply notice that such a scenario would not help in fitting the Spitzer mid-IR points: leaving aside that the envelope would not allow UV radiation to heat the dust, the $\sim 10\%-20\%$ mismatch between either the prism spectrum or IRAC photometry would remain as problematic as the previous fit.

\subsubsection{Blue Rosetta Stone}

In Fig.~\ref{fig:blue-rosetta-sedfit} we explore the red-IR SED fitting of Blue Rosetta, with the same approach as for the Red Rosetta in the previous section.
We start by noticing that fitting the optical to UV continuum in Blue Rosetta is not straightforward. Just as for the SDSS template, it cannot be described by a single powerlaw - it has a steeper index in the blue/UV and a flatter index in the red part of the spectrum. In the case of the SDSS template, the blue slope is interpreted as the mean of the accretion discs of the individual AGN in the stack, subject to some (different) degrees of dust absorption, while the red slope is mostly interpreted in terms of contribution from the host galaxies. In the case of Blue Rosetta the latter interpretation is not so obvious, as there is not much evidence for a prominent, spatially resolved galaxy contributing at these wavelengths, suggesting that the optical radiation is originating from the AGN itself. Here we fit such optical part with a reddened powerlaw ($\beta=-1.85$, $A_V=0.6$), which is disconnected from the blue part of the spectrum (which might originate from nebular emission, as suggested by \citealt{Ji+2024}, and/or a nuclear star cluster, and/or scattered or transmitted continuum); this fit is shown with the orange dashed line. We will discuss later the implications of different fitting models.

The dot-dashed green line shows the torus emission expected from the local type I AGN template, normalized in this case to the 0.75$\mu$m continuum (about the reddest wavelength in the spectrum not affected by emission lines). In this case the expected flux is very close to the observed MIRI-F1800W photometry. 
By performing a full fit, this results into the slightly downscaled AGN1 template, shown with the red solid line. The violet solid line shows the combination of the type I torus template and reddened powerlaw fitting the optical continuum, which in this case reproduces well also the MIRI-F770W photometry and the optical NIRSpec spectrum. We therefore infer that in the case of the Blue Rosetta the observed hot dust emission is pretty much consistent with the expectations from the local templates, especially considering that the latter is the result of multiple sources whose SED scatter by about 0.5 dex around the template.

Fitting the optical spectrum with a simple, unreddened powerlaw, or with a stellar template, would slightly reduce the level of hot dust emission required by the AGN1 template, and it would make the fit worse.

Finally, we also plot the Spitzer IRAC-Ch3 and Ch4 photmetric points (gray circles). These are about 0.1 dex (i.e. 30\%) below the fit and, most importantly, the IRAC-Ch4 photometry is below the MIRI-770W photometry by about the same amount. This highlights either a calibration mismatch between Spitzer and JWST, or variability. In either cases, the same issue may apply to the Red Rosetta Spitzer photometry and may explain the 10\%-20\% discrepancies seen there.

\subsection{Lack of variability}
\label{section-disc-variability}

As outlined in Sect.\ref{sec:timevar},
both Rosetta Stones lack of variability, which is instead typically seen in normal, type I AGN.

In LRDs the lack of variability has been interpreted as the optical radiation being dominated by thermal emission of a pseudo-atmosphere embedding the black hole with a temperature of about 5000K \cite{Liu2025}, responsible for the optical red continuum, while the UV emission might be dominated by the host galaxy or scattered emission \citep{Kokubo+2024}. The former interpretation presents some challenging aspects when compared with various observational evidences (mentioned in the previous sections, and which will be discussed more extensively in a forthcoming paper); however, leaving aside such issues, such an interpretation cannot apply to the Blue Rosetta, which also lacks variability and does not have a red optical continuum that could be interpreted in terms of the putative atmosphere.

A possible solution is that, contrary to previous models, the putative atmosphere is very hot (T$\sim$40,000~K), which \cite{Pacucci2026} suggest may well apply also to LRDs, combined with modest dust reddening.

Yet, without having to invoke new, exotic scenarios, we note that models have expected that the dense BLR clouds (and its outer parts) could be thermalized and emit, themselves thermal continuum, which would have the effect of greatly reducing the variability, if such BLR thermal emission dominates \citep{Baldwin+2004}.
Indeed, continuum emission from the BLR clouds has already been invoked for normal AGN to explain the reduced variability, especially in the red part of the spectrum \citep{Chelouche2019,Korista2001,Netzer2022,Korista2019}. Therefore, it is very possible that the same phenomenon applies to LRDs and LBDs, especially if the covering factor of the BLR is higher and/or the density of the BLR clouds higher.

% , or, more simply, that the emission is more generally associated with super-Eddington accretion, in which case scenarios expect little variability 

Leaving aside the possible interpretations, we note that caution should be adopted when claiming lack of variability. At high redshift, time dilation makes it more challenging to probe variability - indeed, variability is observed when leveraging the time delays of multiply imaged LRD behind lensing clusters \citep{Ji+2025a,Furtak2025,Zhang+2025a}. Additionally, variability is seen when monitoring the EW of the broad Balmer lines \citep{Ji+2025_local,Ji+2025a,Furtak2025,Naidu+2025}, indicating that exploring variability in these systems requires a multi-epoch spectroscopic approach, which is unfortunately available only for very few sources, so far.

\begin{table}
\centering
\caption{Line widths of hydrogen, helium and \oiii lines in GS-3073, based on the purely Gaussian emission line model described in Sect. \ref{section-emission-line-profiles-gs3073}, for both single and total components. For comparison, the results of exponential parametrization (Sect. \ref{section-exp-model}) are also displayed.}
\adjustbox{max width=1.0\textwidth}{
\begin{tabular}{ccc}
\toprule
& \textbf{{Line parameter}} & \textbf{{Value}} [km/s]\\[5pt]
\midrule
\midrule
\multirow{5}{*}{\textbf{Balmer series}} & FWHM$_{Ha,\textrm{narrow},1}$ & $147^{+3}_{-3}$ \\[5pt]
& FWHM$_{\ha,\textrm{narrow},2}$ & $260^{+5}_{-4}$ \\[5pt]
& FWHM$_{\ha,\textrm{broad},1}$ & $1989^{+65}_{-61}$ \\[5pt]
& FWHM$_{\ha,\textrm{broad},2}$ & $4385^{+118}_{-116}$ \\[5pt]
\cline{2-3}
\addlinespace[5pt]
& FWHM$_{\ha,\textrm{broad,tot}}$ & $2453^{+92}_{-77}$ \\[5pt]
\midrule
\multirow{2}{*}{\textbf{\oiii}} & FWHM$_{\oiii\lambda5007,\textrm{outflow},1}$ & $753^{+16}_{-17}$ \\[5pt]
& FWHM$_{\oiii\lambda5007,\textrm{outflow},2}$ & $2991^{+146}_{-138}$ \\[5pt]
\midrule
\multirow{5}{*}{\textbf{\hei}} & FWHM$_{\hei \lambda 7065,\textrm{narrow},1}$ & $185^{+15}_{-13}$ \\[5pt]
& FWHM$_{\hei \lambda 7065,\textrm{narrow},2}$ & $422^{+36}_{-31}$ \\[5pt]
& FWHM$_{\hei \lambda 7065,\textrm{broad},1}$ & $5105^{+382}_{-373}$ \\[5pt]
& FWHM$_{\hei \lambda 7065,\textrm{broad},2}$ & $1761^{+109}_{-95}$ \\[5pt]
\cline{2-3}
\addlinespace[5pt]
& FWHM$_{\hei \lambda 7065,\textrm{broad,tot}}$ & $2052^{+143}_{-143}$ \\[5pt]
\midrule
\textbf{\heii$\lambda 4686$} & FWHM$_{\textrm{\heii4686,~broad}}$ & $7598^{+3}_{-3}$ \\[5pt]
\midrule
\midrule
& FWHM$_{\textrm{\ha,~broad}}$ & $608^{+48}_{-54}$ \\[5pt]
\textbf{Exponential fit} & $W$ & $1036^{+16}_{-15}$ \\[5pt]
& $f_{\textrm{scatt}}$ & $0.91^{+0.02}_{-0.02}$ \\[5pt]
\bottomrule
\end{tabular}}
\label{tab:line_widths_gs3073}
\end{table}

\begin{table}
    \centering
    \caption{Statistics of the three models analysed in this work for broad line emission in GS-3073: single Gaussian, double Gaussian, and exponential.}
    \begin{tabular}{c|c|c|c}
    \toprule
        Model & $\chi ^2_{\rm red}$ & $\chi^2$ & BIC \\
        \midrule
        Single Gauss. &  1.18 & 1589 & 2002 \\
        \midrule
        Double Gauss. & 1.05 & 1406 & 1877 \\
        \midrule
        Exponential & 1.05 & 1406 & 1863 \\
        \bottomrule
    \end{tabular}
    \label{tab:statistics}
\end{table}

\begin{table*}
    \centering
    \begin{tabular}{c|c|c|c}
       \hline 
        Reference source & GN-28074 & GS-3073 & SDSS composite  \\[5pt]
        \hline
         Category & LRDs & LBDs & `Standard' type I  \\[5pt]
         \hline
         Broad line emission & \cmark & \cmark & \cmark  \\[5pt]
         \hline
         Exponential wings in Balmer lines & \cmark & \cmark & \cmark   \\[5pt]
         \hline
         Compactness & \cmark & \cmark & \cmark  \\[5pt]
         \hline
         v-shaped SED & \cmark & \xmark & \xmark \\[5pt]
         \hline
         Excitation diagnostics: &  & & \\
         1. High-ionization lines (\heii 4686) & \xmark & \cmark & \cmark \\
         2. BPT diagram & \xmark & \xmark & \cmark \\
         3. Auroral diagnostic & \cmark & \cmark & \cmark \\[5pt]
         \hline
         Balmer absorption & \cmark & \xmark & \xmark  \\[5pt]
         \hline
         Large (broad c.) Balmer decrement & \cmark & \xmark & \xmark  \\[5pt]
         \hline
         X-ray emission & \xmark & \xmark & \cmark \\[5pt]
         \hline
         Hot dust emission & \cmark & \cmark & \cmark \\[5pt]
         \hline
         Radio emission & \xmark & \xmark & \cmark \\[5pt]
                  \hline
        Variability & \xmark & \xmark & \cmark \\[5pt]
                  \hline
        Overmassive BH$^a$ & \cmark & \cmark & \xmark \\[5pt]
         \hline
    \end{tabular}
    \caption{Comparison summary of observational diagnostics between the two `Rosetta Stones' GN-28074 and GS-3073, and the SDSS quasar stack, where green ticks (red crosses) indicate the presence (absence) of a given feature.\\
    \raggedright{Notes: $^a$BH masses using standard virial relations.}}
    \label{tab:summary-comparison}
\end{table*}

\section{LBDs and LRDs Rosetta Stones: bringing the pieces together}

In this section, we summarise the observational properties of the Blue and Red Rosetta Stones, and put them in the general context of standard type-1 AGN. A summary comparison between the LRD Rosetta, the LBD Rosetta, and the SDSS composite spectrum is provided in Table~\ref{tab:summary-comparison}.

Both Rosetta Stones are broad-line AGN, with broad permitted lines (hydrogen and helium) having widths of thousands of km/s. In both cases, the line profiles are markedly non-Gaussian, with extended wings that are well described by double Gaussians or exponential kernels.
In this context, we remark that even in the `standard' scenario where the broad lines arise from virial motions, there is no theoretical prescription for why the integrated BLR spectrum should have a Gaussian shape \citep[e.g.,][]{Kollatschny2013}. Furthermore, our findings suggest that extended wings are not a prerogative of LRDs, because we find comparable shapes not only in the LBD Rosetta, but even in the SDSS quasar composite and also more generally of several type I AGN \citep[][Trefoloni et al. in prep.]{Laor2006}. For this reason, we suggest that exponential wings should not be considered a defining feature of LRDs. Moreover, the question of the physical origin of the broad wings remains open, whether due to electron or resonant scattering \citetext{see respectively \citealp{Rusakov+2025} and \citealp{Naidu+2025}}, or due to virial motions \citep{Juodzbalis+2025,Juodzbalis+2025b}.
Integral-field spectroscopy of one lensed LRD at $z=7.04$ \citep[Abell2744-QSO1;][]{Furtak2023,Furtak2024} from the BlackTHUNDER programme reveals that the intermediate-velocity component of \ha (a few hundred km/s) is spatially resolved (a few hundred pc), while the broad wings remain spatially unresolved \citep[$< 30$~pc;][]{Juodzbalis+2025b}; such a large spatial mismatch disfavors electron scattering, which predicts that the entire broad line arises from the same compact region. 
The broad component can still be fit with an exponential profile \citep{DEugenio+2025c}, but the resulting black hole mass would be almost two orders of magnitude lower than what measured directly from the spatially resolved velocity field.
Similar studies (of both LRDs and LBDs) are necessary to build a representative picture.

Despite the similarities in their broad-line profiles, the two Rosettas differ sharply in their broad Balmer-line ratios: the LRD Rosetta has a large Balmer decrement \citep[in line with most LRDs;][]{Brooks+2025,Juodzbalis+2024,DeGraaff+2025b}, while the LBD Rosetta has line ratios that are closer to the standard Case-B ratios \citep{Ubler+2023}, similar to the SDSS composite \citep{VandenBerk2001}. Still, several lines of evidence caution against interpreting the Balmer decrement of LBDs as evidence for optically thin regimes \citep[e.g.,][]{Ruff+2012}.

The two Rosetta Stones have remarkably different continuum shape and excitation diagnostics. The Red Rosetta has the v-shaped SED and red optical colours that defines LRDs \citep{Matthee+2024,Perez-Gonzalez+2024,Williams+2024,Hviding+2025}, enabling their efficient photometric selection \citep{Kocevski+2025,Hviding+2025}. In contrast, the Blue Rosetta has a blue power-law continuum very similar to the SDSS composite, with UV and optical slopes which are consistent with the broader population of high-redshift galaxies. This overlap makes LBDs much more difficult to pre-select out of the huge population of star forming galaxies via  photometric colours.

The emission-line diagnostics provide a complex picture. Starting from the BPT diagram \citetext{reported in \citealp{Juodzbalis+2024} and \citealp{Ubler+2023} for the Red and Blue Rosetta, respectively}, neither source occupies the same region as local AGN, due to the low \nii/\ha ratio, which overlaps with low-metallicity starbursts at
$z=5\text{--}7$ \citep[e.g.,][]{Cameron+2023}. In contrast, \heii $\lambda 4686$ is remarkably different: both broad and narrow components are clearly detected in Blue Rosetta, but remain undetected down to stringent upper limits in Red Rosetta. 
Furthermore, in Blue Rosetta we detect both the narrow and broad components, implying that a hard ionizing continuum is reaching not only the BLR, but also the NLR.
The lack of \heii$ \lambda 4686$ in LRDs could be due to an additional obscuring component (such as a dense gas envelope) which efficiently suppresses the highest-energy photons along most lines of sight, while still allowing the broad Balmer lines to escape; such a component could even be the cold, outer layer of the BLR gas, if the covering factor was sufficiently high.

The auroral \oiii diagnostic places the Red Rosetta in the `AGN only' region of the \citet{Mazzolari+2024_newdiag}
diagram in Fig.~\ref{fig:auroral_diagnostic}, while the Blue Rosetta is also consistent with star-formation photoionization. This suggests that Red Rosetta \citep[but also other LRDs;][]{Jones+2025_dormant} likely possess some hard-ionizing photons (X-rays or otherwise) to increase the photon heating per ionization event. Within this context, one possible scenario would be that of low accretion rate AGN SED, whereby plenty of X-ray photons are produced (which would be absorbed by the circumnuclear medium) while the accretion disc blue bump is much weaker, hence reducing the emission of HeII. While we do not have direct measurement of the accretion rate in Red Rosetta, in the other prototypical LRD QSO1, for which the black hole mass is measured and $L_{bol}/L_{Edd}\sim 0.02$ is inferred. Yet, this scenario requires more detailed, quantitative assessment.

The high-temperature hypothesis is particularly interesting because neither Red Rosetta nor Blue Rosetta is X-ray detected, down to stringent upper limits. While the X-ray weakness of Red Rosetta is less severe assuming different bolometric corrections \citep[see, e.g. the extreme scenario proposed by][]{greene+2025}, the object would still be inconsistent with the local relations.
For Red Rosetta, X-ray weakness could plausibly be explained by strong absorption associated with the same medium that imprints the Balmer line absorption and removes the ionizing continuum. For Blue Rosetta, the lack of Balmer absorption disfavours the same explanation; in this case, the co-existence of strong high-ionization lines
\citep{Vanzella+2010,Ubler+2023,Ji+2024} with very weak X-ray emission suggests that ``X-ray weakness'' cannot always be interpreted as simple obscuration, and may instead reflect intrinsic suppression of the corona \citep{Lambrides+2024,Pacucci+2024,Madau+2024,Maiolino+2025,Mazzolari+2025}.

Another strong similarity between the two sources is the presence of MIR emission excess, consistent with thermal dust emission near the dust sublimation temperature. In Blue Rosetta the strength of the mid-IR radiation is comparable with what seen in normal type I AGN; in Red Rosetta it is weaker by a factor of a few, but still clearly detected. This implies that some UV radiation must reach the nuclear dust, therefore the obscuring medium cannot have full covering factor over the central source. In this sense, the presence of hot dust places a strong constraint on scenarios where all the UV emission is absorbed by dense gas: even if significant absorption occurs, there must be optically thin channels through which the AGN powers the hot-dust component.

Yet another puzzling similarity between the two Rosettas is the apparent lack of variability (at least with the photometric data available so far). While the lack of variability in LRDs in the optical part of the spectrum is often interpreted in terms of a warm (T$\sim$5000~K) pseudo-atmosphere, this interpretation cannot apply to the lack of variability in the Blue Rosetta, which shows no evidence for such putative, warm blackbody emission. If the origin of the lack of variability in both Red and Blue Rosettas is the same, then it must be something else. We have suggested that continuum emission from the BLR clouds, already inferred for normal AGN, could potentially explain the reduced or absence of variability in both systems. However, more data are needed to assess the presence/absence of variability, especially with spectroscopic data, as variability in the EW of the emission lines seems to be more common and this would require a revision of all models.

As already mentioned, the two Rosettas differ in the presence of absorption features (most notably Balmer absorption), tracing dense gas. Red Rosetta shows a Balmer break and prominent Balmer absorption, indicating the presence of very dense gas along the line of sight \citep{Juodzbalis+2024}. 
In contrast, for Blue Rosetta there are only indirect lines of evidence, such as weak \civ and (possibly) redshifted \mgii (which, however, could also be interpreted as blended [Fe\,{\sc iv}]$\lambda \lambda 2839,2836$ emission given the presence of many Fe lines in the optical; \citealp{Ji+2024}). This implies that Blue Rosetta offers a more direct view of the accretion disc/BLR -- even though both objects lack X-ray emission. The lack of radio emission in both Rosettas further supports the hypothesis of JWST-selected broad-line AGN occupying a different region of parameter space than luminous quasars.

It has been highlighted by multiple studies that the black hole masses inferred for the population of AGN discovered by JWST at high redshift are overmassive relative to the local BH-stellar mass relation \citep[e.g.][]{Harikane+2023,Maiolino+2024,Juodzbalis+2025,Juodzbalis+2025b,Pacucci2023,Jones2025,Geris2026}. These typically rely on the assumption that the black hole masses can be inferred from the width of the Balmer lines adopting the single-epoch virial relations; these have been questioned in the case of LRDs \citep[e.g.][]{Rusakov+2025}, however for the only LRD for which the black hole mass could be measured directly (QSO1) the resulting value is fully consistent with what inferred from the virial relations \citep{Juodzbalis+2025b}. It is difficult to test this for the broader population of LRDs and it is very possible that for other LRDs the relation between line width and  BH mass deviates from the standard virial relations. Exploring the Black Hole to galaxy mass ratio is more difficult specifically for the Red Rosetta, as no constraints on the stellar mass of the host galaxy are available. Yet, according to \cite{Juodzbalis+2025} it is significantly overmassive relative to the $M_{BH}-\sigma _*$ relation; this is at odds with most of the population of AGN found by JWST that, while being overmassive relative to the $M_{BH}-M_{*}$ relation, are close to the local  $M_{BH}-\sigma_*$ and $M_{BH}-M_{dyn}$ relations. Yet, if the BH mass is significantly overestimated in this specific case, as suggested by \cite{sneppen2026}, then also Red Rosetta would be aligned with such relations. Regarding Blue Rosetta this is also overmassive relative to the local $M_{BH}-M_{star}$ relation \citep{Ubler+2023} and in this case the larger width of the HeII excludes the electron or Balmer scattering scenarios -- even after accounting for bound-free absorption, hence suggesting that the estimation based on the virial relations is probably approximatively correct.

% \fde{Please double check this part in particular}
Taken together, these observations suggest that LRDs and LBDs may share the same central engine (i.e. an accretion disc with a broad-line region and hot dust), but they may differ in the amount and geometry of dense gas, capable of
(i) reddening the UV-optical continuum, generating a v-shaped SED and (ii) imprinting Balmer absorption, and (iii) absorbing the ionizing continuum, such that high-ionization narrow lines are intrinsically weak, and standard ionization diagrams break down. In this scenario, LBDs may even represent lower-obscuration phases of the same AGN phenomenon, while still retaining some of the unique properties (such as X-ray and radio weakness) that set them apart from standard type-1 AGN.
Alternatively, or in addition, LBDs and LDRs may also be characterized by different regimes of accretion (e.g. sub- and super-Eddington) or in terms of properties of the central engine (e.g. presence or absence of a hot corona); in this scenario, the X-ray weakness of LRDs could be due to Compton thick absorption, while it could be due to intrinsic X-ray weakness in the case of LBDs. 
While our work establishes a detailed comparison using two among the most luminous objects as reference, similar studies of larger samples are needed to build a representative picture of the population.

\section{Conclusions}

We have presented a comparative study of the two sources, GS-3073 and GN-28074, taken as the archetypical `Little Blue Dot' and `Little Red Dot', respectively, following the definitions given in section \ref{section-definitions}. 
By highlighting their commonalities and differences, and placing them in relation to luminous quasars from SDSS, we conclude that: 
\begin{itemize}
    \item LRDs are spectroscopically characterized by a v-shaped SED, with optical colours that are typically redder than those of high- and low-redshift broad-line AGN. 
    GN-28074, the Red Rosetta Stone, exhibits precisely these characteristics.
    In contrast, GS-3073, the Blue Rosetta Stone, shows optical and UV slopes consistent with the bulk of $z \gtrsim 2$ broad-line AGN in JADES,
    %with ,
    and also consistent with the bulk of star-forming galaxies.
    This makes the identification of LBDs challenging -- in particular, a simple photometric selection based on optical and UV colours would be ineffective, leaving the  serendipitous identification in spectroscopic surveys the only viable strategy.
    \item Both Red and Blue Rosettas are characterized by broad emission in hydrogen and helium lines, with line widths of $\sim$ few thousands km/s, confirming their AGN nature. 
    If extended to the full LRD and LBD populations, this result suggests that these sources may represent different sub-classes of a common AGN phenomenon, differing from one another, and from standard type I AGN, in the amount and geometry of dense gas, but could differ in terms of accretion regimes (e.g. sub- vs super-Eddington accretion).
    \item Hydrogen lines with exponential profiles are not exclusive to LRDs, given that not only Red Rosetta, but also Blue Rosetta and the SDSS composite -- representative of LRDs, LBDs, and standard type I AGN, respectively -- all exhibit exponential wings in their Balmer lines. 
    While current models of LRDs can account for these wings through electron scattering by free electrons within a dense, partially ionized medium embedding the central source, the physical mechanism and responsible scattering agent remain unclear for LBDs and, more generally, for broad-line AGN; potentially, simple scattering by the partially ionized gas in the BLR clouds can explain this phenomenon \citep{Laor2006}. This indicates that the functional form of hydrogen lines alone is insufficient as a diagnostic of the emission line process itself.
    \item In Blue Rosetta, the correlation of line widths with their ionization potentials is consistent with the BLR stratification seen in normal AGN, where higher-ionization lines originate in the inner regions of the BLR, where gas clouds move faster. 
    \item The detection in the Red Rosetta Stone of the high-ionization line \heii$\lambda 4686$, in both narrow and broad components, implies the presence of a high ionizing flux, further supported by the detection of coronal lines. 
    This radiation is able to reach not only the BLR, but also the NLR. 
    In this scenario, LBDs are expected to exhibit a high intrinsic production of energetic photons --possibly exceeding that of typical type I AGN--, and no significant absorption. 
    In contrast, the absence of \heii4686 in most LRDs, and in particular in Red Rosetta (at least in its narrow component), is possibly due to the presence of the dense gaseous envelope, that efficiently filters high-energy photons, or an intrinsically softer ionizing source.
    % Since such a component is not included in standard type I AGN models, this explains why the excitation diagnostics used to distinguish between AGN and star formation galaxies, both in its classical form of BPT diagram and in terms of \heii4686/\hb versus \nii$\lambda 6584$/\ha, fail in classifying LRDs, an effect further amplified by the fact that these diagrams are mostly calibrated over low-redshift sources exhibiting different physical and chemical properties with respect to higher-redshift sources. 
    \item Both Blue and Red Rosettas are characterized by extreme X-ray weakness. While in the latter this can be ascribed to absorption by the dense gas absorption, as suggested by the presence of deep Balmer absorption, in the former there is no evidence of dense absorbers along the line of sight. Therefore, X-ray weakness could be explained in terms of intrinsic X-ray deficiency, such as the lack of a properly developed hot corona.
    % , but this picture is in contrast with the strong \heii emission, that suggests the presence of an intense hard ionization flux. 
    \item Dust is present in the nuclear regions of both LRDs and LBDs. Furthermore, this dust is hot (T~$\sim 1000$~K), as commonly observed in AGN. Such temperatures imply that the dust is directly exposed to the UV ionizing radiation field, and located close to the sublimation radius. This, in turn, requires that a significant fraction of the AGN-produced UV radiation escapes the central region, in contrast with the simple black-hole star scenario adopted for LRDs. In the Red Rosetta the amount of mid-IR radiation is a factor of 3--6 lower than expected from standard type I AGN templates, but this could be due to ionizing part of the UV radiation being filtered by the neutral absorber, and/or could also be associated with the lower metallicity (hence lower dust-to-gas ratio) in these systems.
    \item Both Red and Blue Rosetta lack clear variability. Reprocessing of the continuum by surrounding gas or by BLR clouds with large covering factor, can potentially explain this phenomenon. 
\end{itemize}

The main results discussed above are summarized in Table \ref{tab:summary-comparison}. 
Our comparative analysis of GS-3073 and GN-28074 probes that, despite sharing a common AGN paradigm, these newly JWST-discovered sources exhibit significant differences both with respect to each other, and to luminous type I AGN commonly observed with ground-based facilities. 
The peculiar properties of LRDs have been the subject of intense investigation in recent months; one of the interpretations proposed recently is the so-called `black-hole star' scenario, according to which the central supermassive black hole is embedded within a dense gaseous envelope, that absorbs most of the emitted ionizing radiation. We have discussed that this scenario presents various observations issues. However, leaving aside those issues, clearly
this framework, does not work for LBDs, as these sources show intermediate characteristics between AGN and LRDs which complicates their identification and classification, especially considering that their analysis so far has been almost entirely overlooked.
The goal of this work is not to provide a definitive interpretation of the nature of LBDs, but rather to highlight their existence as a distinct and peculiar sub-class within the AGN population, characterized by physical properties that differ significantly from those of both classical AGN and LRDs, and that warrant further, more detailed investigation.

\section*{Acknowledgements}

FDE, MB, RM, XJ, IJ, JS, GCJ, SG, AH and LRI acknowledge support by the Science and Technology Facilities Council (STFC), by the ERC through Advanced Grant 695671 ``QUENCH'', and by the UKRI Frontier Research grant RISEandFALL. RM also acknowledges funding from a research professorship from the Royal Society.

The research activities described in this paper were carried out with contribution of the Next Generation EU funds within the National Recovery and Resilience Plan (PNRR), Mission 4 - Education and Research, Component 2 - From Research to Business (M4C2), Investment Line 3.1 - Strengthening and creation of Research Infrastructures, Project IR0000034 – ``STILES - Strengthening the Italian Leadership in ELT and SKA''. 

EB acknowledges financial support from the MIRACLE INAF 2024 GO grant ``A JWST/MIRI MIRACLE: Mid-IR Activity of Circum-nuclear Line Emission'' and the Ricerca Fondamentale INAF 2024 under the project 1.05.24.07.01 MiniGrant RSN1.

CRA thanks the Agencia Estatal de Investigaci\'on of the Ministerio de Ciencia, Innovaci\'on y Universidades (MCIU/AEI) under the grant ``Tracking active galactic nuclei feedback from parsec to kiloparsec scales'', with reference PID2022$-$141105NB$-$I00 and the European Regional Development Fund (ERDF). 

SC, GV, and SZ acknowledge support from the European Union (ERC, WINGS, 101040227).

 AJB acknowledges funding from the “FirstGalaxies” Advanced Grant from the European Research Council (ERC) under the European Union’s Horizon 2020 research and innovation program (Grant agreement No. 789056).

%%%%%%%%%%%%%%%%%%%%%%%%%%%%%%%%%%%%%%%%%%%%%%%%%%

%%%%%%%%%%%%%%%%%%%% REFERENCES %%%%%%%%%%%%%%%%%%

% The best way to enter references is to use BibTeX:

\bibliographystyle{aa}
\bibliography{bib} % if your bibtex file is called example.bib

% Alternatively you could enter them by hand, like this:
% This method is tedious and prone to error if you have lots of references
%\begin{thebibliography}{99}
%\bibitem[\protect\citeauthoryear{Author}{2012}]{Author2012}
%Author A.~N., 2013, Journal of Improbable Astronomy, 1, 1
%\bibitem[\protect\citeauthoryear{Others}{2013}]{Others2013}
%Others S., 2012, Journal of Interesting Stuff, 17, 198
%\end{thebibliography}

%%%%%%%%%%%%%%%%%%%%%%%%%%%%%%%%%%%%%%%%%%%%%%%%%%

%%%%%%%%%%%%%%%%% APPENDICES %%%%%%%%%%%%%%%%%%%%%

\begin{appendix}

\section{Modelling broad Balmer emission in the SDSS composite quasar spectrum}
\label{appendix-ha-fit-in-sdss-composite}

%\mb{Summary of this paragraph: Analysis of different models for Ha emission in the SDSS quasar composite spectrum from \cite{VandenBerk2001}: single broad Gauss, double broad Gauss, exponential. The single broad Gauss model does not work either for this source!}

In this Section, we fit the \ha complex in the quasar composite template from \cite{VandenBerk2001} using the three different models for broad emission analysed in this work: a single broad Gaussian, a broad Gaussian combined with an exponential kernel, and a double broad Gaussian. 
Our results are displayed in Fig. \ref{fig:vandenberk+2001_composite_template_fits}. 
The constraints on the line parameters are the same described in Section \ref{section-emission-line-profiles-gs3073}. 
Given the difficulty in modelling the spectral region around \hb and \oiii due to the strong contamination from \feiiperm, we limit the analysis to the \ha complex only, which we model in its narrow emission with two  components, similarly to GS-3073, and with no outflow. We consider two components also in the nitrogen lines, with kinematics bound to those of \ha. 
We observe that, of the two narrow components, one exhibits $\rm FWHM \ll 500$ km/s, while the other one has $\rm FWHM \sim 1170$ km/s, independently of the model adopted for broad emission. 
In particular, the second narrow component has a width that can also be ascribed to broad emission, or to an outflow, and indeed it is always slightly blueshifted with respect to the first narrow component. 
However, given that we limit our analysis to the fit of few lines, we cannot put further constraints on narrow parameters, and do not delve deeper into their analysis. 

The single broad Gaussian model performs significantly worse than the other two models (Table \ref{tab:statistics-for-sdss-composite}).
The double Gaussian model is the statistically preferred one, but also the exponential model fits the data quite well: therefore, we cannot ascribe exponential profiles in Balmer lines as a prerogative property of LRDs, given that such wings are clearly detected also in the composite spectrum of low-redshift, luminous type I AGN.  In particular, the exponential parameters for the SDSS composite spectrum are not degenerate, and their best values are: $\rm FWHM_{broad} = 2930^{+82}_{-90}$ km/s, $W = 2850 \pm 70$ km/s, and $f_{\rm scatt} = 0.592 \pm 0.014$, implying a broad profile broader than both GN-28074 and GS-3073. 
The single broad Gaussian model has $\rm FWHM_{broad} = 6520 \pm 42$ km/s, but it is clear from Fig. \ref{fig:vandenberk+2001_composite_template_fits} that this model fails in providing a good fit of the line wings. 
The double broad Gaussian model leads to $\rm FWHM_{broad,1} = 3817 \pm 71$ km/s and $\rm FWHM_{broad,2} = 15,444 \pm 200$ km/s. 
We further strengthen these results by carrying out the same \ha fit in other high-resolution local quasar spectra from \cite{dallabonta2025}, and always find the same result: the double broad Gaussian and exponential models perform similarly, and are both significantly better than the single broad Gaussian.

\begin{table}
    \centering
    \caption{Statistics of the three models analysed in this work for broad \ha emission in the SDSS quasar composite from \cite{VandenBerk2001}: single Gaussian, double Gaussian, and exponential.}
    \begin{tabular}{c|c|c|c}
    \toprule
        Model & $\chi ^2_{\rm red}$ & $\chi^2$ & BIC \\
        \midrule
        Single Gauss. &  4.57 & 4041 & 4129 \\
        \midrule
        Double Gauss. & 1.68 & 1485 & 1593 \\
        \midrule
        Exponential & 1.97 & 1735 & 1837 \\
        \bottomrule
    \end{tabular}
    \label{tab:statistics-for-sdss-composite}
\end{table}

\end{appendix}

%%%%%%%%%%%%%%%%%%%%%%%%%%%%%%%%%%%%%%%%%%%%%%%%%%

% \label{lastpage}
\end{document}